\documentclass[twocolumn,showpacs,preprintnumbers,amsmath,amssymb]{revtex4}

\usepackage{graphicx}
\usepackage{dcolumn}
\usepackage{bm}

\newcommand{\dpro}[2]{\langle\hspace{-0.06cm}\langle{#1}|{#2}\rangle
\hspace{-0.06cm}\rangle}
\newcommand{\bra}[1]{\langle{#1}|}
\newcommand{\ket}[1]{|{#1}\rangle}
\newcommand{\bbra}[1]{\langle{\bar{#1}}|}
\newcommand{\bket}[1]{|{\bar{#1}}\rangle}
\newcommand{\dbra}[1]{\langle\hspace{-0.06cm}\langle{#1}|}
\newcommand{\dket}[1]{|{#1}\rangle\hspace{-0.06cm}\rangle}

\newcommand{\dgg}{^{\dagger}}
\newcommand{\Tr}{{\rm Tr}\hspace{0.07cm}}

\newcommand{\im}{{\rm i}}

\newcommand{\bms}{\mbox{\boldmath $S$}}
\newcommand{\bmx}{\mbox{\boldmath $X$}}
\newcommand{\bmy}{\mbox{\boldmath $Y$}}
\newcommand{\bme}{\mbox{\boldmath $E$}}

\newcommand{\bma}{\mbox{\boldmath $A$}}
\newcommand{\bmb}{\mbox{\boldmath $B$}}
\newcommand{\bmh}{\mbox{\boldmath $H$}}
\newcommand{\bmt}{\mbox{\boldmath $T$}}
\newcommand{\bmu}{\mbox{\boldmath $U$}}
\newcommand{\bmp}{\mbox{\boldmath $P$}}

\begin{document}

\preprint{APS/123-QED}

\title{Computational approach to quantum encoder design for purity 
optimization}

\author{Naoki Yamamoto}
 \email{naoki.yamamoto@anu.edu.au}
\affiliation{%
Department of Engineering, Australian National University, 
ACT 0200, Australia 
}%
\author{Maryam Fazel}
 \email{maryam@cds.caltech.edu}
\affiliation{%
Control and Dynamical Systems, California
Institute of Technology, Pasadena, California 91125, USA
}%

\date{\today}

\begin{abstract}

In this paper, we address the problem of designing a quantum encoder 
that maximizes the minimum output purity of a given decohering channel, 
where the minimum is taken over all possible pure inputs. 
This problem is cast as a max-min optimization problem with a rank 
constraint on an appropriately defined matrix variable. 
The problem is computationally very hard because it is non-convex with 
respect to both the objective function (output purity) and the rank 
constraint. 
Despite this difficulty, we provide a tractable computational algorithm 
that produces the exact optimal solution for codespace of dimension two. 
Moreover, this algorithm is easily extended to cover the general class of 
codespaces, in which case the solution is suboptimal in the sense that the 
suboptimized output purity serves as a lower bound of the exact optimal 
purity. 
The algorithm consists of a sequence of semidefinite programmings and can 
be performed easily. 
Two typical quantum error channels are investigated to illustrate the 
effectiveness of our method. 

\end{abstract}

\maketitle


\section{Introduction}

The efficient transmission of quantum states over a noisy channel is
a central subject in quantum information technologies \cite{nielsen}.
The mathematical description of a quantum input-output relation is
as follows. 
Let ${\cal H}$ and ${\cal K}$ be finite-dimensional Hilbert spaces
of an input quantum state and the corresponding output, respectively.
We denote by ${\cal L}({\cal H},{\cal K})$ the set of linear operators
from ${\cal H}$ to ${\cal K}$, and ${\cal S}({\cal H})$ the set of quantum
states on ${\cal H}$.
The Markovian evolution of a quantum state $\rho\in{\cal S}({\cal H})$
through a quantum channel ${\cal A}$ is
typically modeled using the Kraus representation \cite{kraus} as 
\begin{equation}
\label{channel}
    \rho'={\cal A}\rho=\sum_i A_i\rho A_i\dgg,
\end{equation}
where the Kraus operators $A_i\in{\cal L}({\cal H},{\cal K})$ satisfy 
$\sum_i A_i\dgg A_i=I_{{\cal H}}$ with $I_{{\cal H}}$ denoting the identity 
operator on ${\cal H}$. 
The \emph{purity} of a state $\rho$ is defined as $p[\rho]:=\Tr(\rho^2)$, 
which is equal to one if and only if $\rho$ is pure. 
Due to the decoherence caused by ${\cal A}$, a pure input state 
$\rho=\ket{\phi}\bra{\phi}$ may be transmitted to a non-pure output 
$\rho'={\cal A}(\ket{\phi}\bra{\phi})$ with $p[\rho']<1$. 
It is considered that $p[\rho']$ quantifies an intrinsic measure of the 
amount of decoherence induced by the error channel ${\cal A}$. 
In particular, this paper focuses on the {\it optimal purity}:
\begin{equation}
\label{min-purity}
    P({\cal A})
       :=\max_{{\cal C}\subset{\cal H}}\min_{\ket{\phi_c}\in{\cal C}}
            \Tr\big[{\cal A}(\ket{\phi_c}\bra{\phi_c})^2\big],
\end{equation}
where the minimization with respect to the state $\ket{\phi_c}$
takes into account the worst-case scenario of information processing.
The maximization with respect to the {\it codespace} ${\cal C}\subset{\cal H}$
is motivated by the fact that we often have an opportunity to decrease
the effect of decoherence by encoding our information into a 
higher-dimensional space;
this is suggested by the theories of quantum error correction (QEC)
\cite{nielsen,shor,steane,knill} and decoherence-free subspace (DFS)
\cite{lidar1,lidar2,lidar3}.
For example, embedding an input state 
$\ket{\phi}=\phi_1\ket{0}+\phi_2\ket{1}\in{\mathbb C}^2$ into a codespace 
spanned by $\ket{00}$ and $\ket{11}$ through the encoding process 
\begin{equation}
\label{qubit-intro}
   {\mathbb C}^2\ni\ket{\phi}\rightarrow
      \ket{\phi_c}=\phi_1\ket{00}+\phi_2\ket{11}
          \in{\cal C}\subset{\cal H}={\mathbb C}^4
\end{equation}
appears to improve the output purity. 
Clearly, the most desirable situation is the existence of a DFS, i.e., 
a codespace that satisfies $P({\cal A})=1$; but unfortunately this is a 
rare case. 
In this sense, the optimal codespace ${\cal C}$ is regarded as the best 
possible approximation of a DFS.

However, the max-min problem (\ref{min-purity}) is very hard to solve 
because it is non-convex with respect to both ${\cal C}$ and $\ket{\phi_c}$. 
To understand the structure of $P({\cal A})$, in \cite{paolo} Zanardi and 
Lidar considered channel purity for a fixed codespace ${\cal C}$ as 
\begin{equation}
\label{channel-purity}
    P({\cal A}, {\cal C})
       :=\min_{\ket{\phi_c}\in{\cal C}}
            \Tr\big[{\cal A}(\ket{\phi_c}\bra{\phi_c})^2\big], 
\end{equation}
and derived the alternative expression
\[
    P({\cal A}, {\cal C})
             =\min_{\ket{\phi_c}\in{\cal C}}
                    \bra{\phi_c}\otimes\bra{\phi_c}
                          \Omega({\cal A})\ket{\phi_c}\otimes\ket{\phi_c},
\]
where the Hermitian operator $\Omega({\cal A})$ is defined by
\begin{equation}
\label{zanardi-hami}
    \Omega({\cal A}):=\sum_{ij}(A_j\dgg A_i)\otimes(A_i\dgg A_j)
      \in{\cal L}({\cal H}^{\otimes 2}, {\cal H}^{\otimes 2}).
\end{equation}
This expression was used to derive a bound on 
$P({\cal A}, {\cal C})$ in terms of $\Omega({\cal A})$ and ${\cal C}$, 
using techniques to calculate the expectation value of the ``Hamiltonian" 
$\Omega({\cal A})$. 
In the special case where eigenvectors of $\Omega({\cal A})$ are product 
states in a symmetric subspace of ${\cal H}^{\otimes 2}$, analytical 
expressions for $P({\cal A}, {\cal C})$ were obtained. 
However, in general the max-min problem (\ref{min-purity}) does 
not have an analytical solution, leading us to take a computational 
approach.

From a computational point of view, owing to the rapid progress of computers, 
there have been many recent advances with a great potential for solving 
important problems in quantum theory. 
Convex optimization, and in particular semidefinite programming (SDP) 
\cite{boyd1,boyd2}, have proven useful for quantum optimization problems 
such as a test for distinguishing an entangled from a separable 
quantum state \cite{pablo1,pablo2,pablo3,jens,vianna} and a design of optimal 
measurement in linear quantum systems \cite{wiseman}. 
In addition, in \cite{naoki,fletcher,kosut} some quantum error-correction 
problems were solved using SDP, taking advantage of the 
well-known convexity of a set of quantum channels known as the 
{\it Jamiolkowski isomorphism} \cite{jami}.

In this paper, we first use the same convexity property to set up a 
non-convex optimization problem that captures our goal and all the 
constraints. 
Then, we provide an algorithm that computes an {\it exact} local optimal 
solution of the hard non-convex problem (\ref{min-purity}) for the 
codespace of $\dim{\cal C}=2$. 
This implies that the exact global optimal solution of (\ref{min-purity}) 
can be obtained by appropriately choosing an initial condition of the 
algorithm. 
The algorithm is represented by an iterative SDP and is thus computationally 
tractable. 
The derivation of the SDP consists of two stages. 
The first one transforms the constraints to equivalent Linear Matrix 
Inequality (LMI) constraints. 
The key idea used to obtain the LMI in this stage is the {\it Sum-of-Squares} 
characterization of a polynomial constraint \cite{parrilo1,parrilo2,stephen}. 
In the second stage, a non-convex rank constraint of the matrix variable is 
tackled via the {\it log-det} (logarithm of determinant) heuristic 
\cite{fazel1,fazel2,fazel3}. 
Furthermore, we will show an extended version of the above SDP algorithm 
that computes a {\it lower bound} of the optimal purity $P({\cal A})$ 
for the general class of ${\cal C}$.

This paper is organized as follows. 
Section II reviews the Jamiolkowski isomorphism, which is used to formulate 
the optimization problem in Section III. 
The SDP algorithm is presented in Section IV. 
The general case that leads to a suboptimal solution is discussed in 
Section V. 
In Section VI, we examine two typical quantum error channels, the 
{\it bit-flip channel} and the {\it amplitude damping channel}, and 
demonstrate the effectiveness of our method. 
Section VII concludes the paper.

{\it Notation}: 
A Hermitian matrix $X=X\dgg\in{\cal L}({\mathbb C}^n,{\mathbb C}^n)$ 
is {\it positive semidefinite} if 
$\bra{a}X\ket{a}\geq 0,~\forall\ket{a}\in{\mathbb C}^n$; 
the inequality $X\geq 0$ represents the positive semidefiniteness of $X$. 
We use $I_n$ to denote the $n\times n$ identity matrix, which is the same as 
$I_{{\cal H}}$ when $\dim{\cal H}=n$. 
For a matrix $X=(x_{ij})$, the symbols $X^{{\mathsf T}}$ and $X^*$ represent 
the matrix transpose and the elementwise complex conjugate of 
$X$, i.e., $X^{{\mathsf T}}=(x_{ji})$ and 
$X^*=(x_{ij}^*)=(X\dgg)^{{\mathsf T}}$, respectively; 
these rules are applied to any rectangular matrix including column and 
row vectors. 
$\Re(X)$ and $\Im(X)$ denote the real and imaginary part of $X$, respectively, 
i.e., $(\Re(X))_{ij}=(x_{ij}+x_{ij}^*)/2$ and 
$(\Im(X))_{ij}=(x_{ij}-x_{ij}^*)/2\im$.


\section{The Jamiolkowski isomorphism}

The main purpose of this section is to review the following important fact 
known as the Jamiolkowski isomorphism \cite{jami}; 
the set of all finite-dimensional quantum channels has a one-to-one 
correspondence with a convex set of positive semidefinite matrices acting on 
${\cal K}\otimes{\cal H}$. 
This fact can be seen in various ways \cite{choi,naoki,fujiwara,dariano}. 
Here we follow the notations in \cite{naoki,dariano} and obtain two matrix 
representations of a quantum channel, which we later use to set up the 
optimization problem. 
At the end of this section, we present a characterization of
quantum channels that preserve pure states.

We consider a general trace-preserving quantum channel that maps an 
input $\rho\in{\cal S}({\cal H})={\cal S}({\mathbb C}^n)$ to the output
\begin{equation}
\label{kraus-sec2}
     \rho'=\sum_i X_i\rho X_i\dgg
       \in{\cal S}({\cal K})={\cal S}({\mathbb C}^m).
\end{equation}
Let $\{\ket{i}\}_{i=1,\cdots,n}$ and $\{\bket{i}\}_{i=1,\cdots,m}$ 
be orthonormal bases in ${\cal H}$ and ${\cal K}$, respectively. 
Then, any vectors in ${\cal H}^{\otimes 2}$ and ${\cal K}^{\otimes 2}$ are 
expressed as $\dket{\Phi}=\sum_{i,j=1}^{n}\phi_{ij}\ket{i}\otimes\ket{j}$ 
and $\dket{\Phi'}=\sum_{i,j=1}^{m}\phi'_{ij}\bket{i}\otimes\bket{j}$, 
respectively. 
We sometimes use $\ket{i}\ket{j}$ as a short-hand for $\ket{i}\otimes\ket{j}$. 
Let us now define the following two specific vectors: 
\begin{eqnarray}
& & \hspace*{-1em}
\label{1vector}
    \dket{I_{{\cal H}}}:=\sum_{i=1}^{n}\ket{i}\otimes\ket{i}^{*}
        \in{\cal H}^{\otimes 2},
\\ & & \hspace*{-1em}
\label{2vector}
    \dket{I_{{\cal K}}}:=\sum_{i=1}^{m}\bket{i}\otimes\bket{i}^{*}
        \in{\cal K}^{\otimes 2}. 
\end{eqnarray}
These vectors have the property of being independent of the selection of 
orthonormal basis; 
for any two orthonormal bases $\{\ket{a_{i}}\}$ and $\{\ket{b_{i}}\}$ in 
${\cal H}$, we have
\begin{equation}
\label{e-invariant}
    \dket{I_{{\cal H}}}
    =\sum_{i=1}^{n}\ket{a_{i}}\otimes\ket{a_i}^{*}
    =\sum_{i=1}^{n}\ket{b_{i}}\otimes\ket{b_i}^{*}.
\end{equation}
Note that the invariant property (\ref{e-invariant}) is not satisfied if 
$\dket{I_{{\cal H}}}$ is defined without the complex conjugation. 
The vectors (\ref{1vector}) and (\ref{2vector}) are related by 
\begin{equation}
\label{TFDformula}
     (X \otimes I_{{\cal H}})\dket{I_{{\cal H}}}
              =(I_{{\cal K}} \otimes X^{\mathsf T})\dket{I_{{\cal K}}}, 
     ~~\forall X\in{\cal L}({\cal H},{\cal K}). 
\end{equation}
Further, the following equation holds: 
\begin{equation}
\label{trace}
   \dbra{I_{{\cal H}}}(X\otimes I_{{\cal H}})\dket{I_{{\cal H}}}=\Tr X, 
   ~~\forall X\in{\cal L}({\cal H},{\cal H}). 
\end{equation}
We now define a positive semidefinite matrix $\bmx_1$ associated with the 
Kraus operators $X_i\in{\cal L}({\cal H},{\cal K})$ as 
\begin{eqnarray}
& & \hspace*{-1em}
    \bmx_1:=\sum_i(X_i\otimes I_{{\cal H}})\dket{I_{{\cal H}}}
             \dbra{I_{{\cal H}}}(X_i\otimes I_{{\cal H}})\dgg
\nonumber \\ & & \hspace*{5.1em}
             \in{\cal L}({\cal K}\otimes{\cal H},{\cal K}\otimes{\cal H}). 
\nonumber
\end{eqnarray}
Then, the trace-preserving condition $\sum_i X_i\dgg X_i=I_{{\cal H}}$ 
corresponds to $\Tr_{\cal K}\bmx_1=I_{{\cal H}}$, and the quantum channel 
(\ref{kraus-sec2}) is expressed in terms of $\bmx_1$ as 
\begin{equation}
\label{channel-2}
    \rho'=\Tr_{\cal H}\Big[(I_{{\cal K}}\otimes\rho^{\mathsf T})\bmx_1\Big].
\end{equation}
Conversely, it is known that any positive semidefinite matrix 
$\bmx_1\in {\cal L}( {\cal K}\otimes {\cal H},{\cal K}\otimes {\cal H} )$ 
corresponds to a quantum channel with input-output relation given in 
Eq. (\ref{channel-2}). 
That is, there exists a one-to-one correspondence between a quantum channel 
from ${\cal H}$ to ${\cal K}$ and a positive semidefinite matrix on 
${\cal K}\otimes{\cal H}$.

We next introduce another matrix representation of the quantum channel, 
which will be denoted by $\bmx_2$. 
To this end, we define a vector associated with a quantum state 
$\rho\in{\cal S}({\cal H})$ as
\begin{equation}
\label{def-of-vecrho}
    \dket{\rho}:=(\rho \otimes I_{{\cal H}})\dket{I_{{\cal H}}} 
                   \in{\cal H}^{\otimes 2}. 
\end{equation}
The vector $\dket{\rho}$ is obviously in one-to-one correspondence with 
$\rho$. 
In particular, from Eq. (\ref{e-invariant}), the vector representation of 
a pure state $\rho=\ket{a}\bra{a}$ is given by 
\begin{equation}
\label{pure}
    \dket{\rho}=(\ket{a}\bra{a}\otimes I_{{\cal H}})
                    \sum_{i}\ket{i}\otimes\ket{i}^{*}
               =\ket{a}\otimes\ket{a}^{*}. 
\end{equation}
In addition, the purity $p[\rho]=\Tr(\rho^2)$ is simply the squared 
Euclidean norm of $\dket{\rho}$: 
\begin{equation}
\label{purity-vector}
    p[\rho]=\Tr(\rho^2)=\dpro{\rho}{\rho},
\end{equation}
due to Eq. (\ref{trace}). 
Thus, a quantum state $\dket{\rho}$ is pure if and only if
$\dpro{\rho}{\rho}=1$. 
Let us now define $\bmx_2$. 
Multiplying $\dket{I_{{\cal K}}}$ on both sides of Eq. (\ref{kraus-sec2}), 
we have $(\rho'\otimes I_{{\cal K}})\dket{I_{{\cal K}}}
=\sum_i(X_i\rho X_i\dgg\otimes I_{{\cal K}})\dket{I_{{\cal K}}}$,
which is rewritten by 
\begin{eqnarray}
& & \hspace*{-1em}
     (\rho'\otimes I_{{\cal K}})\dket{I_{{\cal K}}}
       =\sum_i(X_i\otimes I_{{\cal K}})(\rho\otimes I_{{\cal K}})
                (I_{{\cal H}}\otimes X^{*}_i)\dket{I_{{\cal H}}}
\nonumber \\ & & \hspace*{5.1em}
       =\sum_i(X_i\otimes X^{*}_i)(\rho\otimes I_{{\cal H}})
                    \dket{I_{{\cal H}}}, 
\nonumber
\end{eqnarray}
because of the property (\ref{TFDformula}). 
Hence, defining the matrix 
\[
     \bmx_2:=\sum_i X_i\otimes X^{*}_i
         \in{\cal L}({\cal H}^{\otimes 2},{\cal K}^{\otimes 2}), 
\]
the quantum channel (\ref{kraus-sec2}) is represented by 
\[
     {\cal H}^{\otimes 2}\ni\dket{\rho}\rightarrow 
      \dket{\rho'}=\bmx_2\dket{\rho}\in{\cal K}^{\otimes 2}. 
\]
The trace-preserving condition is then given by 
\begin{eqnarray}
& & \hspace*{-1em}
    \dbra{I_{{\cal K}}}\bmx_2
     =\sum_i \dbra{I_{{\cal K}}}
                 (X_i\otimes I_{{\cal K}})(I_{{\cal H}}\otimes X^{*}_i)
\nonumber \\ & & \hspace*{2.75em}
     =\sum_i \dbra{I_{{\cal H}}}
           (I_{{\cal H}}\otimes X_i^{{\mathsf T}})(I_{{\cal H}}\otimes X^{*}_i)
     =\dbra{I_{{\cal H}}}. 
\nonumber
\end{eqnarray}
The matrix $\bmx_2$ is related to $\bmx_1$ through the following 
rearrangement rule of the matrix elements: 
\[
     \bbra{i}\bbra{j}^{*}\bmx_2\ket{k}\ket{\ell}^{*} 
      =\bbra{i}\bra{k}^{*}\bmx_1\bket{j}\ket{\ell}^{*}. 
\]
This relation is independent of the selection of $\{\ket{i}\}$ and 
$\{\bket{i}\}$ due to Eq. (\ref{e-invariant}). 
As the rearrangement map is obviously linear and homeomorphic, 
$\bmx_1$ and $\bmx_2$ have a one-to-one correspondence with each other. 
We denote this relation by $\bmx_1=\Phi(\bmx_2)$. 
The above discussion is summarized as follows.

{\bf Lemma 1.} 
Any finite-dimensional quantum channel from ${\cal H}$ to ${\cal K}$ is 
represented by ${\cal H}^{\otimes 2}\ni\dket{\rho}\rightarrow 
\dket{\rho'}=\bmx\dket{\rho}\in{\cal K}^{\otimes 2}$, 
where $\bmx$ is in the set 
\begin{eqnarray}
& & \hspace*{-2.1em}
     {\cal X}({\cal H},{\cal K})
       =\Big\{~\bmx\in{\cal L}({\cal H}^{\otimes 2},{\cal K}^{\otimes 2})
        ~\Big|~
\nonumber \\ & & \hspace*{5em}
      \Phi(\bmx)\geq 0,~
      \dbra{I_{{\cal K}}}\bmx=\dbra{I_{{\cal H}}}~\Big\}.
\nonumber
\end{eqnarray}
The linear transformation $\Phi(\bmx)$ is defined with respect to 
orthonormal bases $\{\ket{i}\}\in{\cal H}$ and $\{\bket{i}\}\in{\cal K}$ as 
\[
     \bbra{i}\bbra{j}^{*}\bmx\ket{k}\ket{\ell}^{*}
      =\bbra{i}\bra{k}^{*}\Phi(\bmx)\bket{j}\ket{\ell}^{*}.
\]

Clearly, ${\cal X}({\cal H},{\cal K})$ is a convex set with dimension 
$m^2n^2-n^2$. 
It should be noted that a cascade connection of two quantum channels 
$\bmx\in{\cal X}({\cal H},{\cal K})$ and 
$\bmy\in{\cal X}({\cal K},{\cal V})$ is simply represented by the 
multiplication of those matrices: 
$\bmy\bmx\in{\cal X}({\cal H},{\cal V})$.

Finally, we provide a characterization of quantum channels that
preserve pure states, i.e., $p[\rho]=p[\rho']=1$, as follows.

{\bf Lemma 2.} 
For a quantum channel $\bmx\in{\cal X}({\cal H},{\cal K})$, the following
three conditions are equivalent.
\begin{eqnarray}
& & \hspace*{-2.1em}
    \mbox{(i) \hspace{0.2cm}
              $\bmx\dket{a}$ is pure for any
              pure state $\dket{a}=\ket{a}\otimes\ket{a}^*$.}
\nonumber \\ & & \hspace*{-2.1em}
    \mbox{(ii)\hspace{0.3cm}}\bmx\dgg\bmx=I_{{\cal H}^{\otimes 2}}=I_{n^2}
\nonumber \\ & & \hspace*{-2.1em}
    \mbox{(iii)~~}{\rm rank}\hspace{0.02cm}\Phi(\bmx)=1
\nonumber
\end{eqnarray}

{\bf Proof.} 
\mbox{(i) $\Leftrightarrow$ (ii)}. 
Condition (ii) immediately implies that $\dket{a'}=\bmx\dket{a}$ 
is pure, since 
$p[a']=\dpro{a'}{a'}=\dbra{a}\bmx\dgg\bmx\dket{a}=\dpro{a}{a}=1$. 
Conversely, as $\bmx$ can be represented by 
$\bmx=\sum_{i=1}^M X_i\otimes X_i^*$, the quantum state 
$\dket{a'}=\bmx\dket{a}$ always satisfies the following relation:
\begin{eqnarray}
\label{lemma3}
& & \hspace*{-1em}
    \dpro{a'}{a'}=\dbra{a}\bmx\dgg\bmx\dket{a}
\nonumber \\ & & \hspace*{2.4em}
     =\bra{a}\bra{a}^*\sum_{i,j}(X_i\dgg\otimes X_i^{{\mathsf T}})
         (X_j\otimes X_j^*)\ket{a}\ket{a}^*
\nonumber \\ & & \hspace*{2.4em}
    =\sum_{i,j}|\bra{a}X_i\dgg X_j\ket{a}|^2
\nonumber \\ & & \hspace*{2.4em}
    \leq\sum_{i,j}\bra{a}X_i\dgg X_i\ket{a}\bra{a}X_j\dgg X_j\ket{a}=1.
\end{eqnarray}
Therefore, the condition $\dpro{a'}{a'}=1$ imposes the equality relation in 
Eq. (\ref{lemma3}). 
Then, $X_i\ket{a}$ is parallel to $X_j\ket{a}$ for all $(i,j)$ and $\ket{a}$, 
indicating that $X_i$ is independent of $i$. 
Thus, $\bmx$ takes the form $\bmx=X\otimes X^*$, where $X$ is defined by 
$X:=\sqrt{M}X_i$. 
Consequently, using the trace-preserving condition $X\dgg X=I_{{\cal H}}$, 
we arrive at $\bmx\dgg\bmx=I_{n^2}$.

\mbox{(ii) $\Leftrightarrow$ (iii)}. 
First, we assume (iii). 
Then, $\Phi(\bmx)$ is written as $\Phi(\bmx)=\dket{x}\dbra{x}$ 
using a vector $\dket{x}\in{\cal K}\otimes{\cal H}$. 
Furthermore, as $\dket{x}$ can be represented by 
$\dket{x}=(X\otimes I_{{\cal H}})\dket{I_{{\cal H}}}$ with a matrix
$X\in{\cal L}({\cal H}\otimes{\cal K})$, we have
$\Phi(\bmx)=(X\otimes I_{{\cal H}})\dket{I_{{\cal H}}}
\dbra{I_{{\cal H}}}(X\otimes I_{{\cal H}})\dgg$, and thus 
$\bmx=X\otimes X^*$ from the definition of $\Phi$. 
This directly yields $\bmx\dgg\bmx=I_{n^2}$ due to 
$X\dgg X=I_{{\cal H}}$. 
We next turn to the proof of \mbox{(ii) $\Rightarrow$ (iii)}. 
Multiplying a pure state 
$\dket{a}=\ket{a}\otimes\ket{a}^*\in{\cal H}^{\otimes 2}$ 
on both sides of $I_{n^2}=\bmx\dgg\bmx$ where $\bmx=\sum_i X_i\otimes X_i^*$, 
we obtain 
\begin{eqnarray}
& & \hspace*{-1em}
    1=\dpro{a}{a}=\dbra{a}\bmx\dgg\bmx\dket{a}
     =\sum_{i,j}|\bra{a}X_i\dgg X_j\ket{a}|^2
\nonumber \\ & & \hspace*{1em}
     \leq\sum_{i,j}\bra{a}X_i\dgg X_i\ket{a}
                   \bra{a}X_j\dgg X_j\ket{a}=1.
\nonumber
\end{eqnarray}
Hence, from the same reason as in the proof of (i) $\Rightarrow$ (ii), 
$\bmx$ must be of the form $\bmx=X\otimes X^*$. 
This implies $\Phi(\bmx)=(X\otimes I_{{\cal H}})\dket{I_{{\cal H}}} 
\dbra{I_{{\cal H}}}(X\otimes I_{{\cal H}})\dgg$ and thus 
${\rm rank}\hspace{0.02cm}\Phi(\bmx)=1$. 
$~\blacksquare$

{\bf Corollary 3.} 
Suppose $\bmx\in{\cal X}({\cal H},{\cal K})$ satisfies 
${\rm rank}\hspace{0.02cm}\Phi(\bmx)=1$. 
Then, the nonzero eigenvalue of $\Phi(\bmx)$ is given by $n=\dim{\cal H}$.

{\bf Proof.} 
From the proof of Lemma 2, we have 
\[
    \Tr\Phi(\bmx)
     =\dbra{I_{{\cal H}}}(X\dgg X\otimes I_{{\cal H}})\dket{I_{{\cal H}}}
     =\dpro{I_{{\cal H}}}{I_{{\cal H}}}
     =n. 
    ~~\blacksquare
\]

According to Lemma 2, the totality of quantum channels that transform 
pure states in ${\cal H}$ to pure in ${\cal K}$ is completely characterized 
by the following non-convex set:
\begin{eqnarray}
& & \hspace*{-2.1em}
     {\cal X}_1({\cal H},{\cal K})
       =\Big\{~\bmx\in{\cal L}({\cal H}^{\otimes 2},{\cal K}^{\otimes 2})
        ~\Big|~ {\rm rank}\hspace{0.02cm}\Phi(\bmx)=1,
\nonumber \\ & & \hspace*{5em}
      \Phi(\bmx)\geq 0,~
      \dbra{I_{{\cal K}}}\bmx=\dbra{I_{{\cal H}}}~\Big\}.
\nonumber
\end{eqnarray}


\section{Optimal encoder design as a matrix optimization problem}

This section is devoted to rewrite the problem (\ref{min-purity}) as an 
encoder-optimization problem, which is further described as a matrix 
optimization problem using the notations introduced in Section II.

First, let us fix the dimension of the codespace ${\cal C}$ to 
$\dim{\cal C}=r$ and represent an element of ${\cal C}$ by 
$\ket{\phi_c}=E\ket{\phi}$ with the input pure state 
$\ket{\phi}\in{\mathbb C}^r$ which contains all information of the sender. 
Here, $E$ is the Kraus operator corresponding to the following encoding 
channel: 
\begin{equation}
\label{encoding}
    {\cal E}:~{\mathbb C}^r \ni \ket{\phi} \rightarrow
         \ket{\phi_c}=E\ket{\phi}\in{\cal C}\subset{\cal H}.
\end{equation}
We set ${\cal H}={\mathbb C}^n$; then, $E$ is an $n\times r$ complex 
matrix satisfying $E\dgg E=I_r$. 
In the example (\ref{qubit-intro}), $\ket{\phi}$ is a qubit and 
$E=\ket{00}\bra{0}+\ket{11}\bra{1}$, i.e., $r=2$ and $n=4$. 
In terms of the above notations, the codespace-optimization problem 
(\ref{min-purity}) is written as 
\begin{eqnarray}
& & \hspace*{-1em}
\label{original}
    P({\cal A})
       =\max_{{\cal E}}\min_{\ket{\phi}\in{\mathbb C}^r}
              P({\cal A},{\cal E},\ket{\phi}),
\nonumber \\ & & \hspace*{-1em}
    P({\cal A},{\cal E},\ket{\phi})
      =\Tr\big[{\cal A}{\cal E}(\ket{\phi}\bra{\phi})^2\big]
\nonumber \\ & & \hspace*{4.4em}
      =\Tr\big[{\cal A}(E\ket{\phi}\bra{\phi}E\dgg)^2\big].
\end{eqnarray}

Next, let us represent the problem using the matrix variable introduced 
in Section II. 
Since the encoding channel ${\cal E}$ obviously preserves pure states, its 
matrix representation $\bme$ is an element of 
${\cal X}_1({\mathbb C}^r,{\mathbb C}^n)$. 
Also, from Eq. (\ref{pure}), the input $\ket{\phi}$ takes the form 
$\dket{\phi}=\ket{\phi}\otimes\ket{\phi}^*$ in the extended space 
$({\mathbb C}^r)^{\otimes 2}$. 
Hence, the output state of the encoder-error process is given by 
$\dket{\rho'}=\bma\bme\ket{\phi}\ket{\phi}^*$, where 
$\bma\in{\cal X}({\mathbb C}^n,{\mathbb C}^n)$ is the matrix representation 
of the error channel ${\cal A}$. 
Then, due to Eq. (\ref{purity-vector}), the output purity is 
\[
     P({\cal A},{\cal E},\ket{\phi})=\dpro{\rho'}{\rho'}
      =\bra{\phi}\bra{\phi}^*\bme\dgg\bma\dgg\bma\bme\ket{\phi}\ket{\phi}^*.
\]
Consequently, the max-min problem (\ref{original}) is written as 
\begin{equation}
\label{redef-problem}
     P({\cal A})=
     \max_{{\bf E}\in{\cal X}_1}\min_{\ket{\phi}\in{\mathbb C}^r}
       \bra{\phi}\bra{\phi}^*\bme\dgg\bma\dgg\bma\bme\ket{\phi}\ket{\phi}^*,
\end{equation}
which is identical to the following ``error-minimization" problem:
\begin{eqnarray}
\label{error-minimization}
& & \hspace*{-1em}
     \min_{{\bf E}, \epsilon}~~~\epsilon, 
\nonumber \\ & & \hspace*{-0.5em}
     \mbox{s.t.}~~
     \bra{\phi}\bra{\phi}^*\bme\dgg\bma\dgg\bma\bme\ket{\phi}\ket{\phi}^*
          \geq 1-\epsilon,~\forall \ket{\phi}\in{\mathbb C}^r, 
\nonumber \\ & & \hspace*{1.7em}
     {\bf E}\in{\cal X}_1({\mathbb C}^r,{\mathbb C}^n), 
\nonumber \\ & & \hspace*{1.7em}
     0\leq\epsilon\leq 1. 
\end{eqnarray}
Note that the optimal purity is related to the minimum error,
$\epsilon_{{\rm opt}}$, by
\[
    P({\cal A})=1-\epsilon_{{\rm opt}}.
\]
%


\section{Exact optimal solution to the purity-optimization problem}

In this section, we provide a systematic and powerful computational 
algorithm that exactly solves the purity-optimization problem when 
$\dim{\cal C}=r=2$. 
The proposed algorithm can easily be extended to cover the general class of 
codespaces of dimension $r\geq 3$, in which case the suboptimized output 
purity gives a lower bound of the optimal purity $P({\cal A})$. 
This result will be discussed in Section~V.

The procedure to derive the algorithm consists of two stages. 
In the first stage, it will be proved that the first constraint in the 
problem (\ref{error-minimization}): 
\begin{equation}
\label{original-constraint}
    \bra{\phi}\bra{\phi}^*\bme\dgg\bma\dgg\bma\bme\ket{\phi}\ket{\phi}^*
          \geq 1-\epsilon,~\forall \ket{\phi}\in{\mathbb C}^2
\end{equation}
can be equivalently transformed to an LMI condition with respect to 
$\bme,~\epsilon$, and an additional variable. 
In the second stage, we will consider a tractable rank-minimization problem 
of the matrix variable that is closely related to the original 
error-minimization problem (\ref{error-minimization}). 
It will be then shown that, under a certain condition, the optimal solution 
of the rank-minimization problem coincides with that of the problem 
(\ref{error-minimization}).


\subsection{The first stage: transformation of the constraint}

To simplify the exposition, we here assume that the input state $\ket{\phi}$ 
is a real-valued qubit, i.e., 
$\ket{\phi}=[x_1, x_2]^{\mathsf T}\in{\mathbb R}^2,~(x_1^2+x_2^2=1)$. 
The general qubit case $\ket{\phi}\in{\mathbb C}^2$ will be discussed 
in Section IV-C using essentially the same idea presented here.

Before considering the transformation of the constraint 
(\ref{original-constraint}), let us further express it only in 
terms of real matrices. 
To this end, we define the following real matrix variable with the size 
$2n^2\times 4$:
\begin{equation}
\label{def-EB}
     \tilde{\bme}
     :=\left[ \begin{array}{c}
             \Re(\bme) \\
             \Im(\bme) \\
       \end{array} \right].
\end{equation}
Then, the output purity is expressed as
\[
     P({\cal A},{\cal E},\ket{\phi})
      =\bra{\phi}\bra{\phi}
       \tilde{\bme}\mbox{}^{{\mathsf T}}\bmp\tilde{\bme}\ket{\phi}\ket{\phi},
\]
where $\bmp$ is a real positive semidefinite matrix defined by 
\[
    \bmp:=
      \left[ \begin{array}{cc}
        \Re(\bma\dgg\bma) & -\Im(\bma\dgg\bma) \\
        \Im(\bma\dgg\bma) & \Re(\bma\dgg\bma) \\
      \end{array} \right]. 
\]
Furthermore, we introduce a matrix
\footnote{
In the absence of $\bmb$, we will need extra real scalar variables 
$\tau_1,\cdots, \tau_5$ in addition to $\bme$ and $\epsilon$ in order to 
obtain an equivalent LMI, whereas our LMI (\ref{lmi}) requires only one 
additional variable $\tau\in{\mathbb R}$. 
}
\begin{equation}
\label{mat-B}
    \bmb:=\left[ \begin{array}{ccc}
              1 & 0 & 0 \\
              0 & 1/\sqrt{2} & 0 \\
              0 & 1/\sqrt{2} & 0 \\
              0 & 0 & 1 \\
          \end{array} \right],
\end{equation}
and define a vector 
\[
     \dket{x}_{B}
         :=\bmb^{{\mathsf T}}\ket{\phi}\ket{\phi}
         =\bmb^{{\mathsf T}} \left[ \begin{array}{c}
              x_1 \\
              x_2 \\
           \end{array} \right]\otimes
           \left[ \begin{array}{c}
              x_1 \\
              x_2 \\
           \end{array} \right]
          =\left[ \begin{array}{c}
              x_1^2 \\
              \sqrt{2}x_1 x_2 \\
              x_2^2 \\
           \end{array} \right].
\]
Note that $\dket{x}_{B}$ is normalized: 
$\mbox{}_{B}\dpro{x}{x}_{B}=1$. 
As a result, from the relation $\ket{\phi}\ket{\phi}=\bmb\dket{x}_{B}$, 
the constraint (\ref{original-constraint}) is expressed as 
\begin{eqnarray}
& & \hspace*{-1em}
\label{preSOS}
   p(x):=
   \mbox{}_{B}\dbra{x}\Big[
     \bmb^{{\mathsf T}}\tilde{\bme}\mbox{}^{{\mathsf T}}\bmp\tilde{\bme}\bmb
         +(\epsilon-1)I_3\Big]\dket{x}_{B} \geq 0,
\nonumber \\ & & \hspace*{14em}
           \forall x_1, x_2\in{\mathbb R}. 
\end{eqnarray}

We are now in the position to describe the transformation. 
The constraint (\ref{preSOS}) indicates that $p(x)$ must be a real 
fourth-order nonnegative polynomial function with respect to the variables 
$(x_1,x_2)$. 
This type of constraint, i.e., the nonnegativity of a polynomial function, 
frequently appears in a wide variety of engineering problems. 
In particular, the following Sum-of-Squares (SOS) characterization of 
non-negative polynomials, first studied by David Hilbert more than a 
century ago, is a fundamental question: 
When does a nonnegative polynomial $p(x)$ have an SOS decomposition 
$p(x)=\sum_i h_i^2(x)$ for some polynomials $h_i(x)$? 
One of the well-known answers to the above question leads us to conclude 
that the nonnegative polynomial $p(x)$ must have an SOS decomposition, 
thereby Eq. (\ref{preSOS}) is equivalently replaced by the following matrix 
inequality: 
\begin{eqnarray}
& & \hspace*{-1em}
\label{eq1}
  \bmb^{{\mathsf T}}\tilde{\bme}\mbox{}^{{\mathsf T}}\bmp\tilde{\bme}\bmb
         +(\epsilon-1)I_3+\tau\bms
\nonumber \\ & & \hspace*{-1em}
  \mbox{}+\bmt_1^{{\mathsf T}}\tilde{\bme}\mbox{}^{{\mathsf T}}
                   \bmp\tilde{\bme}\bmt_2
         +\bmt_2^{{\mathsf T}}\tilde{\bme}\mbox{}^{{\mathsf T}}
                   \bmp\tilde{\bme}\bmt_1
\nonumber \\ & & \hspace*{-1em}
  \mbox{}-\bmt_3^{{\mathsf T}}\tilde{\bme}\mbox{}^{{\mathsf T}}
                   \bmp\tilde{\bme}\bmt_4
         -\bmt_4^{{\mathsf T}}\tilde{\bme}\mbox{}^{{\mathsf T}}
                   \bmp\tilde{\bme}\bmt_3
         \geq 0, 
\end{eqnarray}
where $\tau\in{\mathbb R}$ is an additional optimization variable. 
The proof of Eq. (\ref{eq1}) and the matrices 
$\bmt_1, \bmt_2, \bmt_3, \bmt_4$, and $\bms$ are given in Appendix A. 
The inequality (\ref{eq1}) is transformed to 
\begin{widetext}
\begin{eqnarray}
& & \hspace*{-1em}
\label{eq2}
    \tau\bms+(\epsilon-1)I_3-
    \left[ \begin{array}{c}
             \tilde{\bme}\bmb \\
             \tilde{\bme}\bmt_1 \\
             \tilde{\bme}\bmt_2 \\
             \tilde{\bme}\bmt_3 \\
             \tilde{\bme}\bmt_4 \\
    \end{array} \right]^{{\mathsf T}}
    \left[ \begin{array}{ccccc}
              k I_{2n^2}-\bmp & & & & \\
              & k I_{2n^2} & -\bmp & & \\
              & -\bmp & k I_{2n^2} & & \\
              & & & k I_{2n^2} & \bmp \\
              & & & \bmp & k I_{2n^2} \\
    \end{array} \right]
    \left[ \begin{array}{c}
             \tilde{\bme}\bmb \\
             \tilde{\bme}\bmt_1 \\
             \tilde{\bme}\bmt_2 \\
             \tilde{\bme}\bmt_3 \\
             \tilde{\bme}\bmt_4 \\
    \end{array} \right]
\nonumber \\ & & \hspace*{0em}
   \mbox{}+k\Big[
      \bmb^{{\mathsf T}}\tilde{\bme}\mbox{}^{{\mathsf T}}\tilde{\bme}\bmb
      +\bmt_1^{{\mathsf T}}\tilde{\bme}\mbox{}^{{\mathsf T}}\tilde{\bme}\bmt_1
      +\bmt_2^{{\mathsf T}}\tilde{\bme}\mbox{}^{{\mathsf T}}\tilde{\bme}\bmt_2
      +\bmt_3^{{\mathsf T}}\tilde{\bme}\mbox{}^{{\mathsf T}}\tilde{\bme}\bmt_3
      +\bmt_4^{{\mathsf T}}\tilde{\bme}\mbox{}^{{\mathsf T}}\tilde{\bme}\bmt_4
     \Big]\geq 0,
\end{eqnarray}
\end{widetext}
where the blank spaces in the large matrix denote zero entries. 
The fixed scalar number $k>0$ is selected such that 
\begin{equation}
\label{eq4}
   k I_{2n^2}-\bmp>0
\end{equation}
is satisfied. 
This is equivalent to 
\begin{equation}
\label{eq4-prime}
    \left[ \begin{array}{cc}
        k I_{2n^2} & \bmp \\
        \bmp & k I_{2n^2} \\
     \end{array} \right]>0,~
     \left[ \begin{array}{cc}
        k I_{2n^2} & -\bmp \\
        -\bmp & k I_{2n^2} \\
     \end{array} \right]>0. 
\end{equation}
Then, due to the conditions (\ref{eq4}) and (\ref{eq4-prime}), the large 
matrix in Eq. (\ref{eq2}) is positive definite. 
Moreover, we now see from Lemma 2 that the non-convex rank condition 
${\rm rank}\Phi(\bme)=1$ is equivalent to $\bme\dgg\bme=I_4$, which leads to 
\[
    \tilde{\bme}\mbox{}^{{\mathsf T}}\tilde{\bme}
       =\Re(\bme)^{{\mathsf T}}\Re(\bme)+\Im(\bme)^{{\mathsf T}}\Im(\bme)
       =I_4. 
\]
Thus, the last term in Eq. (\ref{eq2}) is calculated to 
\[
       \bmb^{{\mathsf T}}\bmb
       +\bmt_1^{{\mathsf T}}\bmt_1
       +\bmt_2^{{\mathsf T}}\bmt_2
       +\bmt_3^{{\mathsf T}}\bmt_3
       +\bmt_4^{{\mathsf T}}\bmt_4
         =2I_3.
\]
Finally, the Schur complement (see Appendix B) is used to transform Eq. 
(\ref{eq2}) to
\begin{widetext}
\begin{equation}
\label{lmi}
     \left[ \begin{array}{cccc}
        (k I_{2n^2}-\bmp)^{-1} & & & \tilde{\bme}\bmb \\
         & \left[ \begin{array}{cc}
             k I_{2n^2} & -\bmp \\
             -\bmp & k I_{2n^2} \\
           \end{array} \right]^{-1} & &
         \left[ \begin{array}{c}
             \tilde{\bme}\bmt_1 \\
             \tilde{\bme}\bmt_2 \\
           \end{array} \right] \\
         & &
           \left[ \begin{array}{cc}
             k I_{2n^2} & \bmp \\
             \bmp & k I_{2n^2} \\
           \end{array} \right]^{-1} &
         \left[ \begin{array}{c}
             \tilde{\bme}\bmt_3 \\
             \tilde{\bme}\bmt_4 \\
           \end{array} \right] \\
        \bmb^{{\mathsf T}}\tilde{\bme}\mbox{}^{{\mathsf T}}
           & [\bmt_1^{{\mathsf T}}\tilde{\bme}\mbox{}^{{\mathsf T}}
                   ~\bmt_2^{{\mathsf T}}\tilde{\bme}\mbox{}^{{\mathsf T}}]
           & [\bmt_3^{{\mathsf T}}\tilde{\bme}\mbox{}^{{\mathsf T}}
                   ~\bmt_4^{{\mathsf T}}\tilde{\bme}\mbox{}^{{\mathsf T}}]
           & \tau\bms+(2k+\epsilon-1)I_3
     \end{array} \right]\geq 0,
\end{equation}
\end{widetext}
which is obviously an LMI with respect to the variables $\bme,~\epsilon$, 
and $\tau$. 
As a result, the original problem is equivalently written by 
\begin{eqnarray}
\label{final-form}
& & \hspace*{-1em}
     \min_{{\bf E}, \epsilon, \tau}~~~\epsilon,
\nonumber \\ & & \hspace*{-0.7em}
     \mbox{s.t.}\hspace*{1.3em}
     ({\bf E}, \epsilon, \tau)\in{\cal N}_1, 
\end{eqnarray}
where ${\cal N}_1$ is the following non-convex set: 
\begin{eqnarray}
\label{non-convex-set}
& & \hspace*{-1em}
     {\cal N}_1:=\big\{~(\bme, \epsilon, \tau)~|~
         \Phi(\bme)\geq 0,~
         \dbra{I_n}\bme=\dbra{I_2}, 
\nonumber \\ & & \hspace*{2.5em}
         \mbox{LMI (\ref{lmi})},~
         0\leq\epsilon\leq 1,~
         {\rm rank}\hspace{0.02cm}\Phi(\bme)=1~\big\}. 
\end{eqnarray}
%


\subsection{The second stage: rank-minimization}

Let us consider a closely related problem 
\begin{eqnarray}
\label{final-relaxed}
& & \hspace*{-1em}
     \min_{{\bf E}, \epsilon, \tau}~~~
        {\rm rank}\hspace{0.02cm}\Phi(\bme)+\gamma\epsilon,
\nonumber \\ & & \hspace*{-0.7em}
     \mbox{s.t.}\hspace*{1.3em}
     ({\bf E}, \epsilon, \tau)\in{\cal N}, 
\end{eqnarray}
where ${\cal N}$ is a convex set given by 
\begin{eqnarray}
\label{iterative-SDP-set}
& & \hspace*{-1em}
     {\cal N}:=\big\{~(\bme, \epsilon, \tau)~|~
         \Phi(\bme)\geq 0,~
         \dbra{I_n}\bme=\dbra{I_2}, 
\nonumber \\ & & \hspace*{8em}
         \mbox{LMI (\ref{lmi})},~
         0\leq\epsilon\leq 1~\big\}. 
\end{eqnarray}
The tuning parameter $\gamma>0$ gives the relative weight between 
the two objectives ${\rm rank}\hspace{0.02cm}\Phi(\bme)$ and $\epsilon$. 
This change of the problem is motivated by the fact that we can now apply 
some known heuristic methods for rank minimization problems, one of which 
is discussed below.

\begin{figure}
\includegraphics[scale=0.25]{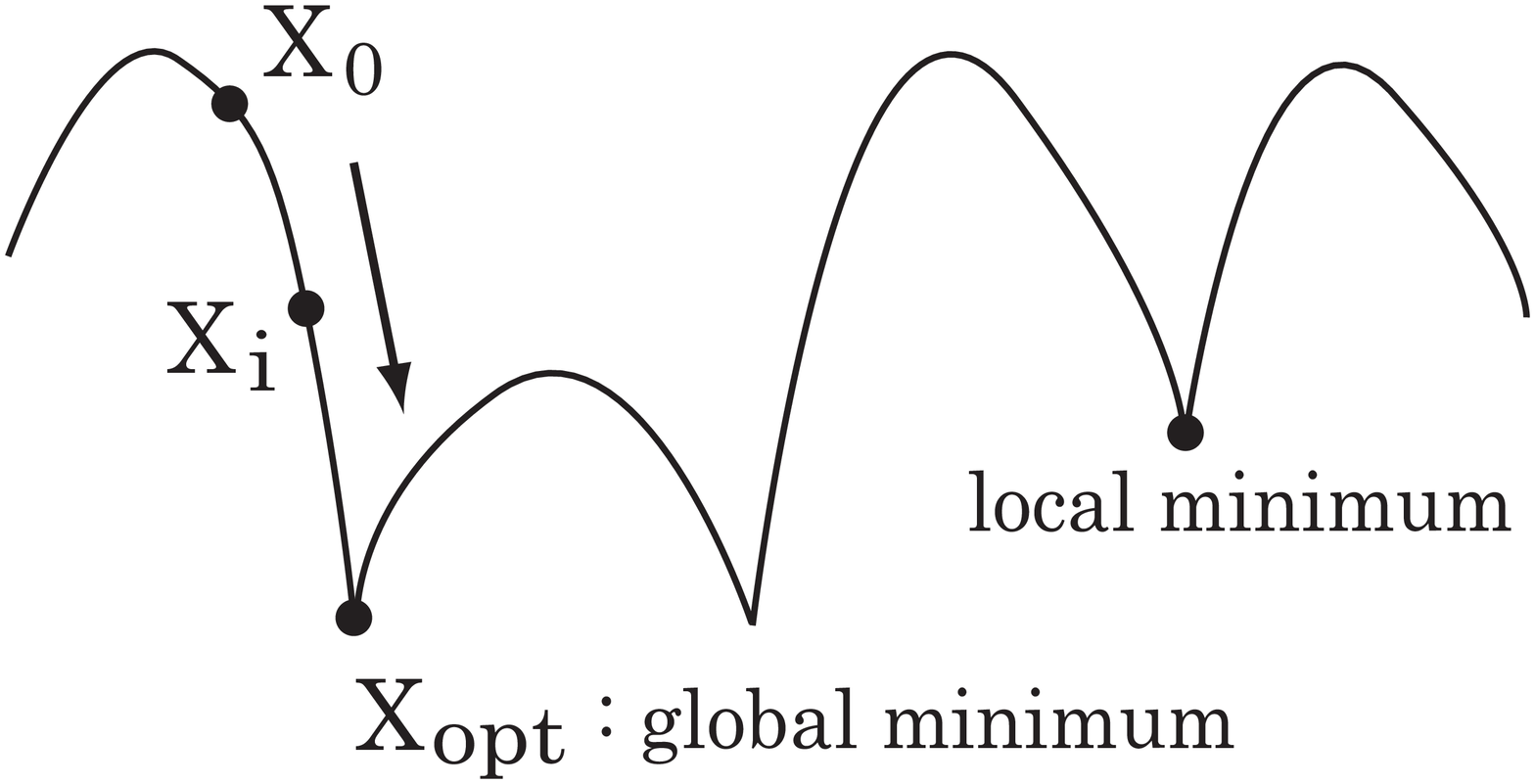}
\caption{\label{logdet}
The log-det function and a convergence of the iteration variable $X_i$. 
}
\end{figure}

The minimization of the rank of a matrix subject to convex constraints 
is a ubiquitous problem in diverse areas of engineering such as control 
theory, system identification, statistics, signal processing, and 
computational geometry \cite{fazel2}. 
The general rank-minimization problem 
\[
    \min~{\rm rank}\hspace{0.05cm}X~~~
    {\rm s.t.}~X\in{\cal M}~\mbox{and}~X\geq 0,
\]
where $X\geq 0$ is the optimization matrix variable and ${\cal M}$ is a 
convex set denoting the constraints, is computationally NP-hard, thus we 
need to rely on heuristics. 
The log-det heuristic introduced and discussed in~\cite{fazel1,fazel2,fazel3} 
provides an attractive approach. 
The heuristic is described as follows: 
The function $\log\det(X+\delta I)$ is used as a {\it smooth surrogate} 
for ${\rm rank}X$ to yield 
\[
    \min~\log\det(X+\delta I)~~~
    {\rm s.t.}~X\in{\cal M}~\mbox{and}~X\geq 0,
\]
where $\delta>0$ is a small regularization constant, and can be chosen to be 
on the order of the eigenvalues we can consider as zero. 
Although the surrogate function $\log\det(X+\delta I)$ is not convex, it is 
smooth on the positive definite cone and can be minimized locally using any 
local minimization method; we here use iterative linearization. 
Let $X_i$ denote the $i$-th iterate of the optimization variable $X$. 
The linearization of $\log\det(X+\delta I)$ around $X_i$ is given by 
\begin{eqnarray}
& & \hspace*{-1em}
\label{local-linear}
     \log\det(X+\delta I)=\log\det(X_i+\delta I)
\nonumber \\ & & \hspace*{6.5em}
     \mbox{}+\Tr\big[ (X_i+\delta I)^{-1}(X-X_i) \big],
\end{eqnarray}
where we have used the fact that $\nabla\log\det X=X^{-1}$ when $X>0$. 
Hence, we can minimize $\log\det(X+\delta I)$ over the constraint set 
${\cal M}$ by iteratively minimizing the local linearization 
(\ref{local-linear}). 
This leads to 
\[
    X_{i+1}={\rm arg}\min_{\hspace{-0.5cm}X\in{\cal M}}
                 \Tr\big[ (X_i+\delta I)^{-1}X \big].
\]
The new optimal point is $X_{i+1}$. 
Since the log-det function is concave in $X$, at each iteration its value 
decreases, and the sequence of the function values generated converges 
to a local minimum of $\log\det(X+\delta I)$. 
This implies that the global optimal solution $X_{\rm opt}$ can be obtained 
by appropriately choosing an initial point $X_0$ (see Fig.~\ref{logdet}).

The above procedure is directly applicable to the case where the objective 
function is replaced by ${\rm rank}X+\gamma\epsilon$ with $\epsilon\in[0,1]$ 
an additional variable and $\gamma>0$ a constant. 
Therefore, the rank-minimization problem (\ref{final-relaxed}) is replaced 
by 
\begin{eqnarray}
\label{log-det-purity}
& & \hspace*{-1em}
     \min_{{\bf E}, \epsilon, \tau}~~~
        \log\det(\Phi(\bme)+\delta I_{2n})+\gamma\epsilon,
\nonumber \\ & & \hspace*{-0.7em}
     \mbox{s.t.}\hspace*{1.3em}
     ({\bf E}, \epsilon, \tau)\in{\cal N}. 
\end{eqnarray}
The local or global optimal solution of this problem is obtained by solving 
the following iterative SDP: 
\begin{eqnarray}
\label{iterative-SDP}
& & \hspace*{-2.8em}
    (\bme_{i+1}, \epsilon_{i+1}, \tau_{i+1})
\nonumber \\ & & \hspace*{-2em}
    =~~
    {\rm arg}\hspace{-0.15cm}
           \min_{\hspace{-0.6cm}({\bf E}, \epsilon, \tau)\in{\cal N}}
        \Big\{ \Tr\big[ (\Phi(\bme_i)
          +\delta I_{2n})^{-1}\Phi(\bme) \big]+\gamma\epsilon \Big\}. 
\end{eqnarray}
Note that the convergence point of this algorithm is very sensitive to an 
initial point $\bme_0$. 
(We do not need to specify $\epsilon_0$ and $\tau_0$, since $\epsilon_i$ and 
$\tau_i$ are not used to calculate 
$(\bme_{i+1}, \epsilon_{i+1}, \tau_{i+1})$.)

We now provide an important theorem that connects the replaced problem 
(\ref{log-det-purity}) with the original problem (\ref{final-form}).

{\bf Theorem 4.} 
If the local (global) optimal solution 
$(\bme_{{\rm opt}}, \epsilon_{{\rm opt}}, \tau_{{\rm opt}})$ of the problem 
(\ref{log-det-purity}) satisfies 
${\rm rank}\hspace{0.02cm}\Phi(\bme_{{\rm opt}})=1$, then it coincides with 
the local (global) optimal solution of the problem (\ref{final-form}).

{\bf Proof.} 
The global optimal solution of (\ref{log-det-purity}) satisfies 
\begin{eqnarray}
\label{theorem4proof}
& & \hspace*{-1em}
    \log\det(\Phi(\bme_{{\rm opt}})+\delta I_{2n})+\gamma\epsilon_{{\rm opt}}
\nonumber \\ & & \hspace*{2em}
    \leq
    \log\det(\Phi(\bme)+\delta I_{2n})+\gamma\epsilon, 
\end{eqnarray}
for all $(\bme, \epsilon, \tau)\in{\cal N}$. 
From the assumption and Corollary~3, the nonzero eigenvalue of 
$\Phi(\bme_{{\rm opt}})$ is $\dim{\mathbb C}^r=2$, which yields 
$\det(\Phi(\bme_{{\rm opt}})+\delta I_{2n})=(2+\delta)\delta^{2n-1}$. 
Also, any $\bme\in{\cal X}_1({\mathbb C}^2, {\mathbb C}^n)$ satisfies 
$\det(\Phi(\bme)+\delta I_{2n})=(2+\delta)\delta^{2n-1}$. 
As a result, Eq. (\ref{theorem4proof}) reduces to 
\[
    \epsilon_{{\rm opt}}\leq\epsilon,~~~
    \forall (\bme, \epsilon, \tau)\in{\cal N}_1, 
\]
where ${\cal N}_1$ is defined in Eq. (\ref{non-convex-set}). 
This implies that $(\bme_{{\rm opt}}, \epsilon_{{\rm opt}}, \tau_{{\rm opt}})$ 
is indeed the global optimal solution of (\ref{final-form}). 
We can prove the same fact for any local optimal solution by considering 
local regions of ${\cal N}$ and ${\cal N}_1$. 
$~\blacksquare$

Clearly, the above theorem can be extended to the general case of $r$. 
Therefore, the optimal (suboptimal) solution of the original problem 
(\ref{error-minimization}) is obtained by equivalently transforming 
(relaxing) it to a problem of the form (\ref{final-form}) and solving 
a related rank-minimization problem via the same heuristic.

Finally, let us discuss choosing an initial point $\bme_0$ of the algorithm 
(\ref{iterative-SDP}) such that the iteration variable 
$(\bme_i, \epsilon_i, \tau_i)$ converges to a local optimal solution 
$(\bme_N, \epsilon_N, \tau_N)$ with ${\rm rank}\hspace{0.02cm}\Phi(\bme_N)=1$. 
We here make the following observation; 
an initial point $\bme_0$ that also satisfies 
${\rm rank}\hspace{0.02cm}\Phi(\bme_0)=1$ might be a good candidate for the 
above requirement to be satisfied. 
From the proof of Lemma 2, this implies 
\begin{equation}
\label{initial-E}
     \bme_0=E_0\otimes E_0^*,~~E_0\dgg E_0=I_{2}. 
\end{equation}
Actually, in many practical cases, we observe that an initial point of the 
form (\ref{initial-E}) converges to a feasible local optimal solution. 
This fact will be seen in Section VI.


\subsection{Exact optimal encoder for general qubit inputs}

We here consider the general qubit input 
$\ket{\phi}=[{\rm e}^{\im\alpha}\cos\beta, \sin\beta]\in{\mathbb C}^2~
(\alpha,\beta\in{\mathbb R})$ and outline the equivalent transformation 
of the constraint (\ref{original-constraint}). 
We first note that the input vector $\ket{\phi}\ket{\phi}^*$ is represented 
in terms of a monomial vector as follows: 
\[
    \ket{\phi}\ket{\phi}^*
    =\left[ \begin{array}{c}
        \cos^2\beta  \\
        {\rm e}^{\im\alpha}\sin\beta\cos\beta \\
        {\rm e}^{-\im\alpha}\sin\beta\cos\beta \\
        \sin^2\beta  \\
     \end{array} \right]
    =\bmu
     \left[ \begin{array}{c}
        x_1^2+x_2^2  \\
        \sqrt{2}x_1x_3 \\
        \sqrt{2}x_2x_3 \\
        x_3^2  \\
     \end{array} \right]
    =:\bmu\dket{x}_U, 
\]
where the real variables $x_1,x_2,x_3\in{\mathbb R}$ are defined as 
\[
    x_1=\cos\beta\cos\alpha,~
    x_2=\cos\beta\sin\alpha,~
    x_3=\sin\beta,
\]
and $\bmu$ is a unitary matrix given by 
\[
    \bmu:=\left[ \begin{array}{cccc}
           1 & & & \\
             & 1/\sqrt{2}~~ & \im/\sqrt{2} & \\
             & 1/\sqrt{2}~~ & -\im/\sqrt{2} & \\
             & & & 1 \\
         \end{array} \right].
\]
Then, similar to the previous case, defining a $2n^2\times 4$ real matrix 
\[
     \tilde{\bme}'
     :=\left[ \begin{array}{c}
             \Re(\bme\bmu) \\
             \Im(\bme\bmu) \\
       \end{array} \right], 
\]
the output purity is expressed by 
\[
     P({\cal A},{\cal E},\ket{\phi})
      =\mbox{}_U\dbra{x}
       \tilde{\bme}'\mbox{}^{{\mathsf T}}\bmp\tilde{\bme}'\dket{x}_U. 
\]
Consequently, the original max-min problem is equal to the minimization of 
$\epsilon\in[0,1]$ subject to the conditions 
$\bme\in{\cal X}_1({\mathbb C}^2,{\mathbb C}^n)$ and 
\begin{eqnarray}
& & \hspace*{-1em}
     p'(x):=\mbox{}_U\dbra{x}\Big[
        \tilde{\bme}'\mbox{}^{{\mathsf T}}\bmp\tilde{\bme}'
             +(\epsilon-1)I_4 \Big]\dket{x}_U\geq 0,
\nonumber \\ & & \hspace*{15.3em}
      \forall x_1,x_2,x_3 \in{\mathbb R}. 
\nonumber
\end{eqnarray}
Since $p'(x)$ is a fourth order homogeneous polynomial with respect 
to the three variables $(x_1,x_2,x_3)$, the Hilbert's lemma (iii) in Eq. 
(\ref{sos-condition}) can be applied; 
the nonnegativity of $p'(x)$ is equivalent to the condition
\[
    \mbox{$p'(x)$ is an SOS with respect to $(x_1,x_2,x_3)$}. 
\]
Then, as the SOS decomposition of $p'(x)$ implies the existence 
of a positive semidefinite matrix $\mbox{\boldmath $Q$}'\geq 0$ 
satisfying $p'(x)=\mbox{}_U\dbra{x}\mbox{\boldmath $Q$}'\dket{x}_U$, 
the matrix 
$\tilde{\bme}'\mbox{}^{{\mathsf T}}\bmp\tilde{\bme}'+(\epsilon-1)I_4$ 
is related to $\mbox{\boldmath $Q$}'$ by 
\begin{eqnarray}
& & \hspace*{-1em}
    \mbox{\boldmath $Q$}'
      =\sum_i \bmt_i'\mbox{}^{{\mathsf T}}\Big[
           \tilde{\bme}'\mbox{}^{{\mathsf T}}\bmp\tilde{\bme}
               +(\epsilon-1)I_4\Big]\bmt_i''
\nonumber \\ & & \hspace*{1em}
  \mbox{}+\sum_i \bmt_i''\mbox{}^{{\mathsf T}}\Big[
      \tilde{\bme}'\mbox{}^{{\mathsf T}}\bmp\tilde{\bme}
          +(\epsilon-1)I_4\Big]\bmt_i'
\nonumber \\ & & \hspace*{1em}
  \mbox{}+\sum_i\tau_i'\bms_i'\geq 0,
\nonumber
\end{eqnarray}
with certain matrices $\bmt_i',\bmt_i'',\bms'_i$, and additional scalar 
variables $\tau_i'\in{\mathbb R}$. 
The above nonlinear matrix inequality with respect to the variables
$\bme,\epsilon$, and $\tau_i'$ is further transformed to an LMI using the 
same technique shown in Section IV-A. 
As before, we then consider the problem of minimizing 
${\rm rank}\Phi(\bme)+\gamma\epsilon$ subject to the LMI obtained above 
and the linear constraints $\Phi(\bme)\geq 0,~\dbra{I_n}\bme=\dbra{I_2}$, 
and $0\leq\epsilon\leq 1$. 
If the optimal solution of this problem satisfies 
${\rm rank}\Phi(\bme_{{\rm opt}})=1$, then it is also the optimal solution 
of the original problem (\ref{error-minimization}) with $r=2$.


\section{Suboptimal solution in higher dimensional code space}

In the general case $r\geq 3$, nonnegativity of a homogeneous polynomial 
no longer implies the existence of its SOS decomposition (this remarkable 
equivalence holds only in the cases (\ref{sos-condition})). 
However, the SOS characterization can still be used as a sufficient 
condition; 
that is, the first constraint in Eq. (\ref{error-minimization}) is 
{\it relaxed} to 
\begin{eqnarray}
& & \hspace*{-1em}
\label{relaxed-sos}
    p''(x)
     :=\dbra{x}\big[\tilde{\bme}''\mbox{}^{{\mathsf T}}
         \bmp\tilde{\bme}''
            +(\epsilon-1)I_{r^2}\big]\dket{x}
\nonumber \\ & & \hspace*{1em}
    \mbox{is an SOS with respect to $(x_1,\ldots,x_{2r-1})$},
\end{eqnarray}
where $\tilde{\bme}''\in{\mathbb R}^{2n^2\times r^2}$ is an appropriately 
defined real matrix variable that is linear to $\bme$, and 
$\dket{x}\in{\mathbb R}^{r^2}$ is an appropriately defined real monomial 
vector of $x_1,\ldots,x_{2r-1}$. 
The SOS condition (\ref{relaxed-sos}) equivalently leads to an LMI as seen 
before, and consequently, we have a problem of the form (\ref{final-form}) 
that can be tackled via the log-det heuristic. 
Note again that Eq. (\ref{relaxed-sos}) is only a sufficient condition 
for the inequality $p''(x)\geq 0$ to be satisfied for all 
$(x_1,\ldots,x_{2r-1})$. 
Thus, any feasible solution $(\tilde{\bme}'',\epsilon)$ satisfying Eq. 
(\ref{relaxed-sos}) is included in the original set of solutions. 
Therefore, the suboptimal error computed from the relaxed problem, 
$\epsilon_{{\rm sub}}$, is always bigger than or equal to the exact optimal 
error $\epsilon_{{\rm opt}}$. 
This indicates that the suboptimal output purity, 
$P_{{\rm sub}}({\cal A})=1-\epsilon_{{\rm sub}}$, gives a lower bound of
the optimal purity: 
\[
   P({\cal A})=1-\epsilon_{\rm opt}
              \geq 1-\epsilon_{\rm sub}=P_{{\rm sub}}({\cal A}). 
\]
An important fact to be noticed is that, as pointed out in \cite{stephen}, 
the gap between the set of nonnegative polynomials and the set of polynomials 
with an SOS decomposition is considered to be small in a practical situation. 
Hence, we expect that $P_{{\rm sub}}({\cal A})$ is a good approximation to 
$P({\cal A})$.


\section{Examples}

\subsection{The bit-flip channel}

The quantum bit-flip channel with flipping probability $p$ is given by
\[
    {\cal S}({\mathbb C}^2)\ni\rho\rightarrow
    {\cal T}_1\rho=p\sigma_{x}\rho\sigma_{x}+q\rho
    \in{\cal S}({\mathbb C}^2),
\]
where $p+q=1$ and $\sigma_{x}=\ket{0}\bra{1}+\ket{1}\bra{0}$. 
We here consider the double bit-flip channel 
${\cal A}_{{\rm bf}}={\cal T}_1^{\otimes 2}$: 
\begin{eqnarray}
& & \hspace*{-1em}
    {\cal S}({\mathbb C}^4)\ni\rho\rightarrow
         \rho'={\cal A}_{{\rm bf}}\rho=\sum_{i=1}^4 A_i\rho A_i\dgg
              \in{\cal S}({\mathbb C}^4),
\nonumber \\ & & \hspace*{0em}
    A_1=p\hspace{0.2em}\sigma_x\otimes\sigma_x,~
    A_2=\sqrt{pq}\hspace{0.2em}\sigma_x\otimes I_2,
\nonumber \\ & & \hspace*{0em}
    A_3=\sqrt{pq}\hspace{0.2em}I_2\otimes\sigma_x,~
    A_4=q\hspace{0.2em}I_2\otimes I_2.
\nonumber
\end{eqnarray}
The matrix form of the double bit-flip channel, 
$\bma_{{\rm bf}}=\sum_{i}A_{i}\otimes A^{*}_{i}
\in{\cal X}({\mathbb C}^{4},{\mathbb C}^{4})$, is represented by
\[
   \bma_{{\rm bf}}=\left[ \begin{array}{cc|cc}
         qA_4 & \sqrt{pq}A_3 & \sqrt{pq}A_2 & pA_1 \\
         \sqrt{pq}A_3 & qA_4 & pA_1 & \sqrt{pq}A_2 \\ \hline
         \sqrt{pq}A_2 & pA_1 & qA_4 & \sqrt{pq}A_3 \\
         pA_1 & \sqrt{pq}A_2 & \sqrt{pq}A_3 & qA_4
        \end{array} \right].
\]
In particular, we set $p=0.1$;
then, for example, $k=2$ satisfies the condition (\ref{eq4}):
$kI_{32}-\bmp>0$.

We here assume that the input is a real-valued qubit: 
$\ket{\phi}\in{\mathbb R}^2$. 
Then, an exact local or global optimal encoder 
$\bme_{{\rm opt}}\in{\cal X}_1({\mathbb R}^2,{\mathbb C}^4)$ is computed by 
the algorithm (\ref{iterative-SDP}) under the condition 
${\rm rank}\hspace{0.02cm}\Phi(\bme_{{\rm opt}})=1$. 
A strong convergence property of $\bme_i$ is observed when the SDP parameters 
are set to $\delta=0.01$ and $\gamma=15$. 
We usually need $90$ iterations of the SDP; 
hence we denote the convergence point by 
$(\bme_{90}, \epsilon_{90}, \tau_{90})$. 
Note again that $\bme_{90}$ must be of the form 
$\bme_{90}=E_{90}\otimes E_{90}^*$ due to the rank condition 
${\rm rank}\hspace{0.02cm}\Phi(\bme_{90})=1$. 
Regarding the initial point $\bme_0$, we follow the idea mentioned in the 
last paragraph of Section IV-B and examine some initial points of the form 
(\ref{initial-E}) to find the global optimal solution.

First, we randomly choose two initial points as 
$\bme_0^{(j)}=E_0^{(j)}\otimes E_0^{(j)*}~(j=1,2)$, where the Kraus operators 
$E_0^{(1)}$ and $E_0^{(2)}$ are given by 
\[
     E_0^{(1)}=
     \frac{1}{\sqrt{10}}
        \left[ \begin{array}{cc}
               2 & 0 \\
               \sqrt{2} & -\sqrt{6} \\
               \sqrt{3} & 1 \\
               1 & \sqrt{3} \\
         \end{array} \right],~~
     E_0^{(2)}=
     \frac{1}{\sqrt{10}}
        \left[ \begin{array}{cc}
               \sqrt{2} & 0 \\
               \sqrt{3} & -\sqrt{6} \\
               1 & \sqrt{2} \\
               2 & \sqrt{2} \\
         \end{array} \right],
\]
respectively. 
Then, the corresponding convergence points are respectively given by 
$\bme_{90}^{(j)}=E_{90}^{(j)}\otimes E_{90}^{(j)*}~(j=1,2)$, where 
\begin{eqnarray}
& & \hspace*{-1em}
     \{~E_{90}^{(1)},~E_{90}^{(2)}~\}
\nonumber \\ & & \hspace*{-1em}
     =\Big\{
         \left[ \begin{array}{cc}
               0.5308 & -0.4672 \\
               0.5308 & -0.4672 \\
               0.4672 & 0.5308 \\
               0.4672 & 0.5308 \\
         \end{array} \right],~~
         \left[ \begin{array}{cc}
               0.5274 & -0.4710 \\
               0.5274 & -0.4710 \\
               0.4710 & 0.5274 \\
               0.4710 & 0.5274 \\
         \end{array} \right]
      \Big\}. 
\nonumber
\end{eqnarray}
In both cases, the convergence value of the error is given by 
$\epsilon_{90}=0.18$. 
In view of the structure of $E_{90}^{(1)}$ and $E_{90}^{(2)}$, we expect 
that the encoder ${\cal E}^{({\rm a})}$: 
\begin{equation}
\label{opt-1}
     \bme^{({\rm a})}=E^{({\rm a})}\otimes E^{({\rm a})*},~~
     E^{({\rm a})}=\frac{1}{\sqrt{2}}
         \left[ \begin{array}{cc}
            \cos\alpha & -\sin\alpha \\
            \cos\alpha & -\sin\alpha \\
            \sin\alpha & \cos\alpha \\
            \sin\alpha & \cos\alpha \\
         \end{array} \right]
\end{equation}
will be a local optimal solution and provide a local minimum of the error, 
$\epsilon^{({\rm a})}=0.18$, for all $\alpha\in[0,2\pi)$. 
In fact, for the input 
$\ket{\phi}=[x_1,~x_2]^{{\mathsf T}}=[\cos\varphi,~\sin\varphi]^{{\mathsf T}}$ 
and the encoder ${\cal E}^{({\rm a})}$, the output purity (\ref{original}) 
is reduced to 
\[
    P({\cal A}_{{\rm bf}},{\cal E}^{({\rm a})},\ket{\phi})
     =1-2pq\big[\cos(2\varphi+2\alpha)\big]^2,
\]
which takes the minimum value 
\[
      P_{{\rm min}}^{{({\rm a})}}
        =\min_{\varphi}
         P({\cal A}_{{\rm bf}},{\cal E}^{({\rm a})},\ket{\phi})
        =1-2pq=0.82.
\]
Hence, as expected above, the local minimum of the error is 
$\epsilon^{({\rm a})}=1-0.82=0.18$. 
This result clarifies that the optimal encoder depends on the worst-case 
input as $\alpha=-\varphi_{{\rm worst}}+n\pi/2$, where $n$ is any integer. 
As a summary, the encoder 
\begin{eqnarray}
& & \hspace*{-1em}
    {\cal E}^{({\rm a})}:~~
       \ket{\phi}
       =[x_1,~x_2]^{{\mathsf T}}
\nonumber \\ & & \hspace*{1em}
     \rightarrow
       \ket{\phi_c}=E^{({\rm a})}\ket{\phi}
        =\frac{1}{\sqrt{2}}
         \left[ \begin{array}{c}
            x_1\cos\alpha-x_2\sin\alpha \\
            x_1\cos\alpha-x_2\sin\alpha \\
            x_1\sin\alpha+x_2\cos\alpha \\
            x_1\sin\alpha+x_2\cos\alpha \\
         \end{array} \right]
\nonumber
\end{eqnarray}
is locally optimal for all $\alpha\in[0,2\pi)$.

We next try following two initial points: 
$\bme_0^{(j)}=E_0^{(j)}\otimes E_0^{(j)*}~(j=3,4)$, where 
\[
     E_0^{(3)}=
     \frac{1}{\sqrt{10}}
        \left[ \begin{array}{cc}
               2 & 0 \\
               \sqrt{2} & \sqrt{6} \\
               \sqrt{3} & -1 \\
               -1 & \sqrt{3} \\
         \end{array} \right],~~
     E_0^{(4)}=
     \frac{1}{\sqrt{10}}
        \left[ \begin{array}{cc}
               \sqrt{3} & -1 \\
               \sqrt{2} & \sqrt{6} \\
               2 & 0 \\
               1 & -\sqrt{3} \\
         \end{array} \right].
\]
Then, the corresponding convergence points are respectively given by 
$\bme_{90}^{(j)}=E_{90}^{(j)}\otimes E_{90}^{(j)*}~(j=3,4)$ where 
\begin{eqnarray}
& & \hspace*{-1em}
     \{~E_{90}^{(3)},~E_{90}^{(4)}~\}
\nonumber \\ & & \hspace*{-1em}
     =\Big\{
         \left[ \begin{array}{cc}
               0.6935 & -0.1377 \\
               0.1377 & 0.6935 \\
               0.6935 & -0.1377 \\
               0.1377 & 0.6935 \\
         \end{array} \right],~~
         \left[ \begin{array}{cc}
               0.4636 & -0.5341 \\
               0.5341 & 0.4636 \\
               0.5341 & 0.4636 \\
               0.4636 & -0.5341 \\
         \end{array} \right]
      \Big\}. 
\nonumber
\end{eqnarray}
Although they have a similar structure, there is a large gap between
the corresponding convergence values of the error $\epsilon$:
\[
     \epsilon_{90}^{(3)}=0.18,~~
     \epsilon_{90}^{(4)}=0.2952.
\]
The structure of $E_{90}^{(3)}$ and $E_{90}^{(4)}$ suggests that the encoders 
$\bme^{(\mu)}=E^{(\mu)}\otimes E^{(\mu)}\mbox{}^*$ $(\mu={\rm b},{\rm c})$ 
with 
\begin{eqnarray}
& & \hspace*{-1em}
     \{~E^{({\rm b})},~E^{({\rm c})}~\}
\nonumber \\ & & \hspace*{-1em}
     =\Big\{~\frac{1}{\sqrt{2}}
         \left[ \begin{array}{cc}
            \cos\alpha & -\sin\alpha \\
            \sin\alpha & \cos\alpha \\
            \cos\alpha & -\sin\alpha \\
            \sin\alpha & \cos\alpha \\
         \end{array} \right],~~
     \frac{1}{\sqrt{2}}
         \left[ \begin{array}{cc}
            \cos\alpha & -\sin\alpha \\
            \sin\alpha & \cos\alpha \\
            \sin\alpha & \cos\alpha \\
            \cos\alpha & -\sin\alpha \\
         \end{array} \right]~\Big\} 
\nonumber
\end{eqnarray}
are locally optimal for all $\alpha\in[0,2\pi)$. 
Actually, the output purity (\ref{original}) with the above encoders and 
the input $\ket{\phi}=[\cos\varphi,~\sin\varphi]^{{\mathsf T}}$ are 
respectively calculated as 
\begin{eqnarray}
& & \hspace*{-1em}
    P({\cal A}_{{\rm bf}},{\cal E}^{({\rm b})},\ket{\phi})
     =1-2pq[\cos(2\varphi+2\alpha)]^2,
\nonumber \\ & & \hspace*{-1em}
    P({\cal A}_{{\rm bf}},{\cal E}^{({\rm c})},\ket{\phi})
     =1-4pq(p^2+q^2)[\cos(2\varphi+2\alpha)]^2. 
\nonumber
\end{eqnarray}
Thus, their minimum values are 
\begin{eqnarray}
& & \hspace*{-1em}
     P_{{\rm min}}^{({\rm b})}=1-2pq=0.82,
\nonumber \\ & & \hspace*{-1em}
     P_{{\rm min}}^{({\rm c})}=1-4pq(p^2+q^2)=0.7048,
\nonumber
\end{eqnarray}
irrespective of $\alpha$.
The minimums are attained when $\cos(2\varphi+2\alpha)=\pm 1$, 
as in the case of ${\cal E}^{({\rm a})}$. 
We also see the following inequality: 
\[
    P_{{\rm min}}^{({\rm b})}-P_{{\rm min}}^{({\rm c})}=2pq(1-2p)^2\geq 0. 
\]
Therefore, the encoders ${\cal E}^{({\rm a})}$ and ${\cal E}^{({\rm b})}$ 
achieve the same local minimum of the error, whereas 
${\cal E}^{({\rm c})}$ is inferior to those channels for all $p$.

Combining the entire set of investigations presented above with other 
numerical results that were omitted for brevity, we maintain that 
$\epsilon_{{\rm opt}}=0.18$ is the global minimum and that the optimal 
purity is thus given by 
\[
     P({\cal A}_{{\rm bf}})=1-\epsilon_{{\rm opt}}=0.82. 
\]
The solutions ${\cal E}^{({\rm a})}$ and ${\cal E}^{({\rm b})}$ 
are typical optimal encoders that yield the above optimal purity. 
\\
\\
\indent
{\bf Remark 1.}~
In the Kraus representation, the output state is given by 
$\rho'=\sum_i A_i E\ket{\phi}\bra{\phi}E\dgg A_i\dgg$. 
Intuitively, in order for the purity of $\rho'$ to have a large value, 
the encoder $E$ should be chosen so that the vectors 
$\{A_i E\ket{\phi}\}$ are close to each other. 
Actually, if all of them are parallel, the output state is pure. 
In this sense, ${\cal E}^{({\rm a})}$ is a physically reasonable encoder 
because the vectors $A_1 E^{({\rm a})}\ket{\phi}$ and 
$A_3 E^{({\rm a})}\ket{\phi}$ are parallel to 
$A_2 E^{({\rm a})}\ket{\phi}$ and 
$A_4 E^{({\rm a})}\ket{\phi}$, respectively. 
The encoders $E^{({\rm b})}$ and $E^{({\rm c})}$ also satisfy such relations. 
In contrast, if we choose $E=\ket{00}\bra{0}+\ket{11}\bra{1}$ in Eq. 
(\ref{qubit-intro}), the four vectors $A_i E\ket{\phi}~(i=1,\ldots,4)$ 
differ from each other and span a linear space of dimension $4$. 
This is indeed a bad encoder since the minimum output purity in this case 
is calculated as $p[\rho']=(p^2+q^2)^2\approx 0.67$, which is clearly less 
than the optimal purity $P({\cal A}_{{\rm bf}})=0.82$.

{\bf Remark 2.}~
We again maintain that $\bme_0$ satisfying ${\rm rank}\Phi(\bme_0)=1$ is a 
good initial point. 
Actually, within our investigation, we have observed that such an initial 
point always converges to a rank-one solution by appropriately choosing 
the SDP parameters $\delta$ and $\gamma$. 
However, for initial points with the rank more than one, it is easy to 
find a bad example of $\bme_0$ such that ${\rm rank}\Phi(\bme_{90})=1$ is 
not achieved for any $\delta$ and $\gamma$. 
For instance, if we choose $\Phi(\bme_0)=(1/4)I_8$, then $\bme_i$ always 
converges to a solution satisfying ${\rm rank}\Phi(\bme_{90})=2$. 
Another reason of the emphasis is based on the following observation. 
Once we obtain a rank-one solution using an initial point with the rank more 
than one, then we always have found a rank-one initial point that converges 
to the same solution. 
In other words, it is considered that any rank-one solution is available 
by choosing a rank-one initial point appropriately. 
For example, $\bme_i$ starting from 
$\bme_0=0.5\bme_0^{(1)}+0.3\bme_0^{(2)}+0.2\bme_0^{(3)}$ converges into 
a rank-one solution of the form (\ref{opt-1}).


\subsection{The amplitude damping channel}

The amplitude damping channel describes the dissipation of a quantum 
state into equilibrium due to coupling with its environment. 
The Kraus representation of the channel is given by 
\[
    {\cal S}({\mathbb C}^2)\ni\rho\rightarrow 
    {\cal T}_2\rho=H_1\rho H_1\dgg + H_2\rho H_2\dgg 
    \in{\cal S}({\mathbb C}^2), 
\]
where 
\[
    H_1=\left[ \begin{array}{cc} 
         1 & 0 \\
         0 & \sqrt{p} \\
        \end{array} \right],~
   H_2=\left[ \begin{array}{cc}
         0 & \sqrt{1-p} \\
         0 & 0 \\
       \end{array} \right]. 
\]
The parameter $p\in(0,1)$ represents the rate of dissipation. 
We consider the double amplitude damping channel 
${\cal A}_{{\rm ad}}={\cal T}_2^{\otimes 2}$: 
\begin{eqnarray}
& & \hspace*{-1em}
    {\cal S}({\mathbb C}^4)\ni\rho\rightarrow
         \rho'={\cal A}_{{\rm ad}}\rho=\sum_{i=1}^4 A_i\rho A_i\dgg
              \in{\cal S}({\mathbb C}^4),
\nonumber \\ & & \hspace*{0em}
    A_1=H_1\otimes H_1,~
    A_2=H_1\otimes H_2,
\nonumber \\ & & \hspace*{0em}
    A_3=H_2\otimes H_1,~
    A_4=H_2\otimes H_2.
\nonumber
\end{eqnarray}
The matrix form of the channel, 
$\bma_{{\rm ad}}=\sum_{i}A_{i}\otimes A^{*}_{i}
\in{\cal X}({\mathbb C}^{4},{\mathbb C}^{4})$, is given by 
\[
   \bma_{{\rm ad}}=\left[ \begin{array}{cc|cc}
         A_1 & \sqrt{1-p}A_2 & \sqrt{1-p}A_3 & (1-p)A_4 \\
         O_4   & \sqrt{p}A_1 & O_4 & \sqrt{p(1-p)}A_3 \\ \hline
         O_4   & O_4 & \sqrt{p}A_1 & \sqrt{p(1-p)}A_2 \\
         O_4   & O_4 & O_4 & pA_1
        \end{array} \right],
\]
where $O_4$ denotes the $4\times 4$ zero matrix. 
In particular, we consider the case of $p=0.1$ and set $k=4$, which leads 
to $kI_{32}-\bmp>0$.

Our goal is to obtain the optimal encoder under the condition 
$\ket{\phi}\in{\mathbb R}^2$, in which case 
$\bme_{{\rm opt}}\in{\cal X}_1({\mathbb R}^2,{\mathbb C}^4)$. 
The iteration variable $\bme_i$ of the algorithm (\ref{iterative-SDP}) is 
initialized to $\bme_0$ of the form (\ref{initial-E}), and the SDP parameters 
are set to $\delta=0.01$ and $\gamma=6.1$. 
In order to find a rank-one convergence point, we usually need $500$ 
iterations of the SDP; 
we thus denote the convergence point by 
$(\bme_{500}, \epsilon_{500}, \tau_{500})$.

First, let us take the initial points $\bme_0^{(1)}$ and $\bme_0^{(3)}$, 
which have appeared in the bit-flip channel case. 
Then, the corresponding convergence points are respectively given by 
$\bme_{500}^{(j)}=E_{500}^{(j)}\otimes E_{500}^{(j)*}~(j=5,6)$ 
with the Kraus operators 
\begin{eqnarray}
& & \hspace*{-1em}
     \{~E_{500}^{(5)},~E_{500}^{(6)}~\}
\nonumber \\ & & \hspace*{-1em}
     =\Big\{
         \left[ \begin{array}{cc}
               0.6555 & 0.2503 \\
               0.4174 & -0.9045 \\
               0.5741 & 0.2822 \\
               0.2583 & 0.1989 \\
         \end{array} \right],~~
         \left[ \begin{array}{cc}
               0.5803 & -0.2273 \\
               0.4668 & 0.8745 \\
               0.5568 & -0.2843 \\
               0.3679 & -0.3209 \\
         \end{array} \right]
      \Big\}. 
\nonumber
\end{eqnarray}
In both cases, the convergence value of the error is given by 
$\epsilon_{500}=0.18$. 
Unlike the case of bit-flip channel, the above solutions do not have 
a simple structure of the matrix entries, which is highly important for 
a physical realization of encoding process. 
To obtain a simple solution, let us carry out the algorithm with an initial 
point that has a specific matrix form itself. 
As a typical example, we consider the following initial point: 
\[
     \bme_0^{({\rm d})}=E_0^{({\rm d})}\otimes E_0^{({\rm d})*},~~
     E_0^{({\rm d})}=\left[ \begin{array}{cc}
               \cos\alpha & 0 \\
               0          & \cos\beta \\
               \sin\alpha & 0 \\
               0          & \sin\beta \\
         \end{array} \right].
\]
Then, for any $\alpha\in[0,2\pi)$ and $\beta\in(0,\pi/2)$, $\bme_i$ 
converges to 
\[
     \bme_{500}^{({\rm d})}
         =E_{500}^{({\rm d})}\otimes E_{500}^{({\rm d})*},~~
     E_{500}^{({\rm d})}=\left[ \begin{array}{cc}
               \cos\alpha & 0 \\
               0          & 1 \\
               \sin\alpha & 0 \\
               0          & 0 \\
         \end{array} \right], 
\]
with $\epsilon_{500}^{({\rm d})}=0.18$. 
This encoder is locally optimal for all $\alpha\in [0,2\pi)$. 
Actually, the output purity (\ref{original}) for the encoder-error 
process ${\cal A}_{{\rm ad}}{\cal E}_{500}^{({\rm d})}$ with the input 
$\ket{\phi}=[x_1,~x_2]^{{\mathsf T}}$ is calculated as 
\begin{equation}
\label{ad-pure1}
     P({\cal A}_{{\rm ad}},{\cal E}_{500}^{({\rm d})},\ket{\phi})
     =1-2p(1-p)(x_1^2 \sin^2\alpha+x_2^2)^2, 
\end{equation}
and thus, its minimum is $P_{{\rm min}}^{({\rm d})}=1-2p(1-p)=0.82$ 
when $\ket{\phi}=[0,1]^{{\mathsf T}}$ irrespective of $\alpha$.

We also observe the following similar convergence:
\[
     E_0^{({\rm e})}=\left[ \begin{array}{cc}
               \cos\alpha & 0 \\
               \sin\alpha & 0 \\
               0          & \cos\beta \\
               0          & \sin\beta \\
         \end{array} \right]~\rightarrow~
     E_{500}^{({\rm e})}=\left[ \begin{array}{cc}
               \cos\alpha & 0 \\
               \sin\alpha & 0 \\
               0          & 1 \\
               0          & 0 \\
         \end{array} \right],
\]
with $\epsilon_{500}^{({\rm e})}=0.18$ for all $\alpha\in[0,2\pi)$ and 
$\beta\in(0,\pi/2)$. 
The output purity 
$P({\cal A}_{{\rm ad}},{\cal E}_{500}^{({\rm e})},\ket{\phi})$ 
has the same form as Eq. (\ref{ad-pure1}), thus the encoder 
${\cal E}_{500}^{({\rm e})}$ is also locally optimal for all 
$\alpha\in[0,2\pi)$.

Finally, let us choose an initial point of the form
\begin{equation}
\label{ad-initial-2}
     \bme_0^{({\rm f})}
         =E_0^{({\rm f})}\otimes E_0^{({\rm f})*},~~
     E_0^{({\rm f})}=\left[ \begin{array}{cc}
               \cos\alpha & 0 \\
               0          & \cos\beta \\
               0          & \sin\beta \\
               \sin\alpha & 0 \\
         \end{array} \right]. 
\end{equation}
We then observe a somewhat complicated convergence depending on 
$(\alpha,\beta)$ as follows. 
When $\alpha$ takes a small number, e.g., $\alpha=0.2$ 
(any $\beta$ can be taken), the algorithm does not cause a variation in 
$\bme_i$, and only $\epsilon_i$ changes into $0.18$. 
That is, we obtain the local optimal solution 
\[
   \bme_{500}^{({\rm f_1})} 
   =E_{500}^{({\rm f_1})}\otimes E_{500}^{({\rm f_1})}\mbox{}^*,~~
     E_{500}^{({\rm f_1})}=\left[ \begin{array}{cc}
               \cos\alpha & 0 \\
               0          & \cos\beta \\
               0          & \sin\beta \\
               \sin\alpha & 0 \\
         \end{array} \right]. 
\]
On the other hand, when $\alpha\approx\pi/2$, another type of convergence 
occurs. 
For example when choosing $\alpha=1.3$, $\bme_i$ converges to 
\[
    \bme_{500}^{({\rm f_2})}
    =E_{500}^{({\rm f_2})}\otimes E_{500}^{({\rm f_2})}\mbox{}^*,~~
    E_{500}^{({\rm f_2})}=\left[ \begin{array}{cc}
               0.6893 & 0 \\
               0    & \cos\beta \\
               0    & \sin\beta \\
               0.7245 & 0 \\
         \end{array} \right], 
\]
with $\epsilon_{500}^{({\rm f_2})}=0.18$. 
To further understand this complex structure of the solution, we provide 
an analytical investigation of the output purity 
$P({\cal A}_{{\rm ad}}, {\cal E}_0^{({\rm f})},\ket{\phi})$ in Appendix~C. 
However, we reemphasize that a lucid advantage of our method to search an 
optimal solution is that it does not require any analytic examination on 
the max-min optimization problem of the output purity, which is in general 
extremely hard.

Based on the above investigations, we maintain that 
$\epsilon_{{\rm opt}}=0.18$ is the global minimum and that 
${\cal E}_{500}^{(\mu)}~(\mu={\rm d},{\rm e},{\rm f_1},{\rm f_2})$ 
are the optimal encoders. 
Therefore, the optimal purity is given by 
\[
    P({\cal A}_{{\rm ad}})=1-\epsilon_{{\rm opt}}=0.82. 
\]
%


\section{Conclusion}

In this paper, we presented a tractable computational algorithm for 
designing a quantum encoder that maximizes the worst-case output purity of 
a given decohering channel over all possible pure inputs. 
We cast the problem as a max-min optimization problem (minimization 
over all pure inputs, and maximization over all pure state preserving 
encoders). 
Although this problem is computationally very hard to solve due to the 
non-convexity property, our algorithm computes the exact optimal solution 
for codespace of dimension two. 
Moreover, we showed an extended version of the above algorithm that computes 
a lower bound of the optimal purity for the general class of codespaces.

We believe that the proposed computational approach provides a powerful 
method that is also applicable to other problems in quantum encoding and 
fault-tolerant quantum information transmissions. 
For example, following the same techniques presented in this paper, we can 
prove that a quantum error correction problem with the minimum fidelity 
criterion considered in \cite{knill,naoki} is transformed or relaxed to a 
convex optimization problem systematically; 
we are then able to obtain the optimal or suboptimal solution using SDP. 
This result will be reported soon.


\begin{acknowledgements}

We wish to thank P. Parrilo for pointing out the SOS characterization. 
NY would like to acknowledge stimulating discussions with S. Hara and 
H. Siahaan. 
MF thanks M. Yanagisawa for helpful discussions. 
This work was supported in part by the Grants-in-Aid for JSPS fellows 
No.06693.

\end{acknowledgements}


\appendix


\section{Proof of Eq. (\ref{eq1})}

Let us consider a real polynomial function $p(x)$ in $n$ variables 
$x=(x_1,\ldots,x_n)$ of the form: 
\[
    p(x)=\sum_{k}c_k x_1^{k_1}\cdots x_n^{k_n},~~c_k\in{\mathbb R}, 
\]
where the sum is over $n$-tuples $k=(k_1,\ldots,k_n)$ satisfying 
$\sum_{i=1}^{n}k_i=m$. 
This function is called the {\it homogeneous polynomial} of degree $m$ in 
$n$ variables. 
A homogeneous polynomial satisfies 
$p(\lambda x_1,\ldots,\lambda x_n)=\lambda^m p(x_1,\ldots,x_n)$. 
We now state the famous Hilbert's theorem. 
Let $P_{n,m}$ be the set of nonnegative homogeneous polynomials of degree 
$m$ in $n$ variables. 
Let $\Sigma_{n,m}$ be the set of homogeneous polynomials $p(x)$ that has an 
SOS decomposition $p(x)=\sum_i h_i(x)^2$, where $h_i(x)$ are homogeneous 
polynomials of degree $m/2$. 
Then, $P_{n,m}=\Sigma_{n,m}$ holds only in the following cases: 
\begin{equation}
\label{sos-condition}
   \mbox{(i)}~n=2~~~~\mbox{(ii)}~m=2~~~~\mbox{(iii)}~n=3,~m=4. 
\end{equation}
For more detailed description on this problem, see \cite{reznick}.

Now, Eq. (\ref{preSOS}) has the following form: 
\begin{equation}
\label{appB-1}
    p(x)=[x_1^2~\sqrt{2}x_1x_2~x_2^2]\bmh
           \left[ \begin{array}{c}
           x_1^2 \\
           \sqrt{2}x_1x_2 \\
           x_2^2
           \end{array} \right]\geq 0,~~
           \forall x_1, x_2\in{\mathbb R},
\end{equation}
where $\bmh=(h_{ij})$ is a real $3\times 3$ symmetric matrix. 
The function $p(x)$ is a homogeneous polynomial with respect to two 
variables $x_1$ and $x_2$ (and degree $m=4$).
Therefore, from the Hilbert's formula (i) in Eq. (\ref{sos-condition}), 
the constraint (\ref{appB-1}) is equivalent to the condition 
\[
   \mbox{$p(x)$ is an SOS with respect to $x_1$ and $x_2$}.
\]
Moreover, it can be shown that the existence of an SOS decomposition is
equivalent to the existence of a positive semidefinite matrix
$\mbox{\boldmath $Q$}=(q_{ij})\geq 0$ such that
\begin{equation}
\label{appB-2}
  p(x)=z(x)^{{\mathsf T}}\mbox{\boldmath $Q$}z(x),
\end{equation}
where $z(x)$ is a vector of monomials of degree equal to 
${\rm deg}(p)/2=2$. 
Comparing Eq. (\ref{appB-2}) with (\ref{appB-1}), we set 
$z(x)=[x_1^2,~\sqrt{2}x_1x_2,~x_2^2]^{{\mathsf T}}$. 
Then, the equality 
$z(x)^{{\mathsf T}}\bmh z(x)=z(x)^{{\mathsf T}}\mbox{\boldmath $Q$}z(x)$ 
yields 
\begin{eqnarray}
& & \hspace*{-1em}
   h_{11}=q_{11},~
   h_{12}=q_{12},~~
   h_{13}+h_{22}=q_{13}+q_{22},
\nonumber \\ & & \hspace*{-1em}
   h_{23}=q_{23},~
   h_{33}=q_{33},
\nonumber
\end{eqnarray}
which leads to
\[
   \mbox{\boldmath $Q$}
    =\left[ \begin{array}{ccc}
      h_{11} & h_{12} & q_{13} \\
      h_{12} & h_{22}+h_{13}-q_{13} & h_{23} \\
      q_{13} & h_{23} & h_{33} \\
     \end{array} \right]\geq 0.
\]
As a result, Eq. (\ref{appB-1}) is equivalent to the following matrix 
inequality:
\[
    \left[ \begin{array}{ccc}
      h_{11} & h_{12} & 0 \\
      h_{12} & h_{22}+h_{13} & h_{23} \\
      0 & h_{23} & h_{33} \\
    \end{array} \right]
    +\tau\left[ \begin{array}{ccc}
        0 & 0 & 1 \\
        0 & -1 & 0 \\
        1 & 0 & 0 \\
      \end{array} \right]\geq 0,
\]
where $\tau:=q_{13}\in{\mathbb R}$ is an additional optimization variable.
The above inequality can be expressed as 
\[
   \bmh
     +\bms_1^{{\mathsf T}}\bmh\bms_2
     +\bms_2^{{\mathsf T}}\bmh\bms_1
     -\bms_3^{{\mathsf T}}\bmh\bms_4
     -\bms_4^{{\mathsf T}}\bmh\bms_3
     +\tau\bms\geq 0,
\]
where
\begin{eqnarray}
& & \hspace*{-1em}
   \bms_1
      =\left[ \begin{array}{ccc}
        0 & 1/\sqrt{2} & 0 \\
        0 & 0 & 0 \\
        0 & 0 & 0 \\
       \end{array} \right],~
   \bms_2
      =\left[ \begin{array}{ccc}
        0 & 0 & 0 \\
        0 & 0 & 0 \\
        0 & 1/\sqrt{2} & 0 \\
       \end{array} \right],
\nonumber \\ & & \hspace*{-1em}
   \bms_3
      =\left[ \begin{array}{ccc}
        1 & 0 & 0 \\
        0 & 0 & 0 \\
        0 & 0 & 0 \\
       \end{array} \right],~
   \bms_4
      =\left[ \begin{array}{ccc}
        0 & 0 & 0 \\
        0 & 0 & 0 \\
        0 & 0 & 1 \\
       \end{array} \right],~
   \bms=\left[ \begin{array}{ccc}
        0 & 0 & 1 \\
        0 & -1 & 0 \\
        1 & 0 & 0 \\
       \end{array} \right].
\nonumber
\end{eqnarray}
From the above discussion, the constraint (\ref{preSOS}) is equivalently 
transformed to 
\begin{eqnarray}
& & \hspace*{-1em}
    \bmb^{{\mathsf T}}\tilde{\bme}\mbox{}^{{\mathsf T}}\bmp\tilde{\bme}\bmb
          +(\epsilon-1)I_3+\tau\bms
\nonumber \\ & & \hspace*{0em}
    \mbox{}
    +\bms_1^{{\mathsf T}}\Big[
         \bmb^{{\mathsf T}}\tilde{\bme}\mbox{}^{{\mathsf T}}
            \bmp\tilde{\bme}\bmb
               +(\epsilon-1)I_3\Big]\bms_2
\nonumber \\ & & \hspace*{0em}
    \mbox{}
    +\bms_2^{{\mathsf T}}\Big[
         \bmb^{{\mathsf T}}\tilde{\bme}\mbox{}^{{\mathsf T}}
            \bmp\tilde{\bme}\bmb
               +(\epsilon-1)I_3\Big]\bms_1
\nonumber \\ & & \hspace*{0em}
    \mbox{}
    -\bms_3^{{\mathsf T}}\Big[
         \bmb^{{\mathsf T}}\tilde{\bme}\mbox{}^{{\mathsf T}}
            \bmp\tilde{\bme}\bmb
               +(\epsilon-1)I_3\Big]\bms_4
\nonumber \\ & & \hspace*{0em}
    \mbox{}
    -\bms_4^{{\mathsf T}}\Big[
         \bmb^{{\mathsf T}}\tilde{\bme}\mbox{}^{{\mathsf T}}
            \bmp\tilde{\bme}\bmb
               +(\epsilon-1)I_3\Big]\bms_3
    \geq 0.
\nonumber
\end{eqnarray}
As $\bms_1^{{\mathsf T}}\bms_2=O$ and $\bms_3^{{\mathsf T}}\bms_4=O$, 
we obtain Eq. (\ref{eq1}):
\begin{eqnarray}
& & \hspace*{-1em}
   \bmb^{{\mathsf T}}\tilde{\bme}\mbox{}^{{\mathsf T}}\bmp\tilde{\bme}\bmb
        +(\epsilon-1)I_3+\tau\bms
\nonumber \\ & & \hspace*{-1em}
   \mbox{}+\bmt_1^{{\mathsf T}}\tilde{\bme}\mbox{}^{{\mathsf T}}
      \bmp\tilde{\bme}\bmt_2
          +\bmt_2^{{\mathsf T}}\tilde{\bme}\mbox{}^{{\mathsf T}}
             \bmp\tilde{\bme}\bmt_1
\nonumber \\ & & \hspace*{-1em}
   \mbox{}-\bmt_3^{{\mathsf T}}\tilde{\bme}\mbox{}^{{\mathsf T}}
      \bmp\tilde{\bme}\bmt_4
          -\bmt_4^{{\mathsf T}}\tilde{\bme}\mbox{}^{{\mathsf T}}
             \bmp\tilde{\bme}\bmt_3
          \geq 0,
\nonumber
\end{eqnarray}
where the matrices $\bmt_i~(i=1,\ldots,4)$ are defined as 
\[
    \bmt_1:=\bmb\bms_1,~\bmt_2:=\bmb\bms_2,~
    \bmt_3:=\bmb\bms_3,~\bmt_4:=\bmb\bms_4.
\]
%


\section{The Schur complement}

The Schur complement is a powerful tool that transforms a convex but 
nonlinear constraint with respect to matrix variables into an equivalent LMI. 
Its derivation is very easy; 
assuming $A>0$, we have a matrix equation of the form: 
\begin{eqnarray}
& & \hspace*{-2em}
     \left[ \begin{array}{cc}
         I & O \\
         -B\dgg A^{-1} & I \\
     \end{array} \right]
     \left[ \begin{array}{cc}
         A & B \\
         B\dgg & C \\
     \end{array} \right]
     \left[ \begin{array}{cc}
         I & -A^{-1}B \\
         O & I \\
     \end{array} \right]
\nonumber \\ & & \hspace*{1em}
     =\left[ \begin{array}{cc}
         A & O \\
         O & C-B\dgg A^{-1}B \\
     \end{array} \right].
\nonumber
\end{eqnarray}
Hence, the following relation holds: 
\[
     \left[ \begin{array}{cc}
         A & B \\
         B\dgg & C \\
     \end{array} \right]\geq 0 ~\Leftrightarrow~
     \left\{ \begin{array}{cc}
         \hspace{-1.9cm} A>0 \\
         C-B\dgg A^{-1}B\geq 0. \\
     \end{array} \right.
\]
This is termed the Schur complement. 
In order to see the usefulness, let us consider a 
nonlinear constraint of a matrix variable $X$: $I-X\dgg X\geq 0$. 
The Schur complement states that the constraint is equivalent to 
\[
    \left[ \begin{array}{cc}
         I & X \\
         X\dgg & I \\
    \end{array} \right]\geq 0,
\]
which is obviously an LMI.


\section{An analytic investigation of the purity-optimization problem}

We here give an observation on the purity-optimization problem 
where the error channel is ${\cal A}_{{\rm ad}}$ and the encoder is 
${\cal E}_0^{({\rm f})}\rho=E_0^{({\rm f})}\rho E_0^{({\rm f})}\mbox{}^*$ 
with $E_0^{({\rm f})}$ given in Eq. (\ref{ad-initial-2}). 
The output purity 
$P({\cal A}_{{\rm ad}},{\cal E}_0^{({\rm f})},\ket{\phi})
=\Tr[{\cal A}_{{\rm ad}}{\cal E}_0^{({\rm f})}(\ket{\phi}\bra{\phi})^2]$
with the input $\ket{\phi}=[x_1,~x_2]^{{\mathsf T}}\in{\mathbb R}^2$ 
is then calculated to 
\begin{eqnarray}
& & \hspace*{-1.5em}
   P(\alpha,\beta,x_1)
      =1-2pq\Big[ (1+\sin 2\alpha\sin 2\beta-2pq\sin^4\alpha)x_1^4
\nonumber \\ & & \hspace*{4.55em}
     \mbox{}-(1+\cos 2\alpha+\sin 2\alpha\sin 2\beta)x_1^2+1 \Big].
\nonumber
\end{eqnarray}
First, let us consider the case where $\alpha$ takes a small number. 
Especially when $\alpha=0$, $P(0,\beta,x_1)=1-2pq(x_1^2-1)^2$ is a concave 
function with respect to $x_1$. 
Thus, the minimum is given by $P_{{\rm min}}=P(0,\beta,0)=1-2pq$ at $x_1=0$. 
This fact is still true for $\alpha\approx 0$; 
the function $P(\alpha,\beta,x_1)$ is concave and takes the minimum $1-2pq$ 
at $x_1=0$ without respect to the values of $\alpha$ and $\beta$. 
This is the reason why $\alpha$ and $\beta$ do not have specific optimal 
values and the iterative SDP initialized with $\alpha\approx 0$ does not 
renew these parameters. 
On the other hand, when $\alpha=\pi/2$, the output purity becomes
\[
   P(\pi/2,\beta,x_1)=1-2pq\Big[ (p^2+q^2)x_1^4+1 \Big],
\]
which obviously takes the minimum at $x_1=1$. 
Moreover, for $\alpha\approx\pi/2$ the function $P(\alpha,\beta,x_1)$ is 
still concave and takes the minimum 
$P(\alpha,\beta,1)=1+4p^2 q^2 \sin^2\alpha (\sin^2\alpha-1/pq)$. 
Unlike the case of $\alpha\approx 0$, this function must be further 
maximized with respect to $\alpha$. 
For this reason, there is a specific optimal value of $\alpha$, whereas 
$\beta$ does not affect the optimality.


\begin{thebibliography}{99}

\bibitem{nielsen}
M. A. Nielsen and I. L. Chuang,
{\it Quantum Computation and Quantum Information},
Cambridge University Press, Cambridge (2000).

\bibitem{kraus}
K. Kraus,
{\it States, Effects and Operations, Fundamental Notions of
Quantum Theory} (Academic, Berlin, 1983).

\bibitem{shor}
P. Shor,
Phys. Rev. A 52, 2493 (1995).

\bibitem{steane}
A. M. Steane,
Phys. Rev. Lett. 77, 793 (1996).

\bibitem{knill}
E. Knill and R. Laflamme,
Phys. Rev. A 55, 900 (1997).

\bibitem{lidar1}
D. A. Lidar, I. L. Chuang, and K. B. Whaley,
Phys. Rev. Lett. 81, 2594 (1998).

\bibitem{lidar2}
D. A. Lidar, D. Bacon, and K. B. Whaley,
Phys. Rev. Lett. 82, 4556 (1999).

\bibitem{lidar3}
A. Shabani and D. A. Lidar,
Phys. Rev. A 72, 042303 (2005). 

\bibitem{paolo}
P. Zanardi and D. A. Lidar,
Phys. Rev. A 70, 012315 (2004).

\bibitem{boyd1}
L. Vandenberghe and S. Boyd, SIAM Review 38, 49 (1996). 

\bibitem{boyd2}
S. Boyd, L. El Ghaoui, E. Feron, and V. Balakrishnan,
{\it Linear matrix inequalities in systems and control theory},
(SIAM, Philadelphia, 1994).

\bibitem{pablo1}
A. C. Doherty, P. A. Parrilo, and F. M. Spedalieri,
Phys. Rev. Lett. 88, 187904 (2002).

\bibitem{pablo2}
A. C. Doherty, P. A. Parrilo, and F. M. Spedalieri,
Phys. Rev. A 69, 022308 (2004).

\bibitem{pablo3}
A. C. Doherty, P. A. Parrilo, and F. M. Spedalieri,
Phys. Rev. A 71, 032333 (2005).

\bibitem{jens}
J. Eisert, P. Hyllus, O. Guhne, and M. Curty, 
Phys. Rev. A 70, 062317 (2004). 

\bibitem{vianna}
R. O. Vianna and A. C. Doherty, 
Phys. Rev. A 74, 052306 (2006). 

\bibitem{wiseman}
H. M. Wiseman and A. C. Doherty,
Phys. Rev. Lett. 94, 070405 (2005).

\bibitem{naoki}
N. Yamamoto, S. Hara and, K. Tsumura, 
Phys. Rev. A 71, 022322 (2005).

\bibitem{fletcher}
A. S. Fletcher, P. W. Shor, and M. Z. Win, 
Phys. Rev. A 75, 012338 (2007).

\bibitem{kosut}
R. L. Kosut and D. A. Lidar, 
e-print quant-ph/0606078 (2006). 

\bibitem{jami}
A. Jamiolkowski,
Rep. Math. Phys. 3, 275 (1972).

\bibitem{parrilo1}
P. Parrilo,
{\it Structured semidefinite programs and semialgebraic geometry
methods in robustness and optimization}, 
(Ph.D thesis, California Institute of Technology, Pasadena, CA, 2000). 

\bibitem{parrilo2}
P. Parrilo, 
Mathematical Programming Ser. B, 96, 2 293/320 (2003).

\bibitem{stephen}
S. Prajna, A. Papachristodoulou, P. Seiler, and P. Parrilo,
{\it Positive Polynomials in Control}, 273/292, 
(Springer, 2005).

\bibitem{fazel1}
M. Fazel, H. Hindi, and S. P. Boyd, 
Proceedings of American Control Conference, June 2003. 

\bibitem{fazel2}
M. Fazel, 
{\it Matrix rank minimization with applications}, 
(Ph.D thesis, Stanford University, Stanford, CA, 2002).

\bibitem{fazel3}
M. Fazel, H. Hindi, and S. P. Boyd, 
Proceedings of American Control Conference, June 2001. 

\bibitem{choi}
M. D. Choi,
Linear Algebr. Appl. 10, 285 (1975).

\bibitem{fujiwara}
A. Fujiwara and P. Algoet,
Phys. Rev. A 59, 3290 (1999).

\bibitem{dariano}
G. M. D'Ariano and P. Lo Presti,
Phys. Rev. A 64, 042308 (2001).

\bibitem{reznick}
B. Reznick,
{\it Contemporary Mathematics}, 253, 251/272, 
(AMS, 2000).



\end{thebibliography}
\end{document}